\documentclass[12pt]{article}

\usepackage{a4wide}
\usepackage{amsmath}
\usepackage{amsthm}
\usepackage{amsfonts}
\usepackage{amssymb}
\usepackage{enumerate}
\usepackage{bbm}
\usepackage{cite}
\usepackage{graphicx}

\newcommand{\N}{\mathbbm{N}}

\newcommand{\R}{\mathbbm{R}}

\newcommand{\eg}{e.g.\ }
\newcommand{\ie}{i.e.\ }
\newcommand{\e}{\mathrm{e}}
\newcommand{\ii}{\mathrm{i}}
\newcommand{\order}[1]{{O({#1})}}
\newcommand{\ol}[1]{\omega_L \!\! \left({#1}\right)}

\newtheorem{theorem}{Theorem}[section]
\newtheorem{lemma}[theorem]{Lemma}

\numberwithin{equation}{section}

\begin{document}
\thispagestyle{empty}
\vspace*{-80pt} 
{ \hspace*{\fill} Preprint-KUL-TF-2001/22} 
\vspace{60pt} 
\begin{center} 
{\LARGE  Fluctuations in the Bose Gas with Attractive \\[5pt]Boundary Conditions
  }
 \\[25pt]  
 
{\large  
    J.~Lauwers\footnote{Bursaal KUL FLOF-10408}\footnotetext{Email: {\tt joris.lauwers@fys.kuleuven.ac.be}},
	A.~Verbeure\footnote{Email: {\tt andre.verbeure@fys.kuleuven.ac.be}}
    } \\[25pt]   
Instituut voor Theoretische Fysica,  
Katholieke Universiteit Leuven,  
Celestijnenlaan~200D,  
B-3001 Leuven, Belgium\\[25pt]
{November 2001}\\[25pt]
\end{center} 
\begin{abstract}\noindent
In this paper limiting distribution functions of field and density fluctuations
are explicitly and rigorously computed for the different phases of the Bose gas. 
Several Gaussian and non-Gaussian distribution functions are obtained and
the dependence on boundary conditions is explicitly derived. The model under 
consideration is the free Bose gas subjected to attractive boundary conditions, 
such boundary conditions yield a gap in the spectrum.  
The presence of a spectral gap and the method of the coupled thermodynamic limits 
are the new aspects of this work, leading to new scaling exponents and new 
fluctuation distribution functions.   
\\[15pt]
\begin{center}
{\bf  Keywords}
\end{center}
Quantum Fluctuations, Distribution Functions,
Bose-Einstein Condensation, Boundary Conditions.
\end{abstract}

\newpage


\section{Introduction}
Normal and critical fluctuations in the ideal Bose gas with Dirichlet or
periodic boundary conditions are explicitly studied at different levels of
rigour in \cite{wreszinski:1974, ziff:1977, buffet:1983,  fannes:1983, 
nachtergaele:1985, tuyls:1995,broidioi:1996, angelescu:1996, michoel:1999b,
zagrebnov:2001}. Condensation occurs only at dimensions $\nu \geq 3$.
It turns out that the behaviour of the energy gap of the finite volume spectrum
as a function of the volume is determining for the degree of abnormality of the
fluctuations in the condensate regime. Typical is that the limit spectrum has no
energy gap. The ideal Bose gas with attractive boundary conditions
\cite{robinson:1976,landau:1979} on the other hand is quite different in nature. Condensation is
possible in all space dimensions, and the spectrum has a finite energy gap in
the thermodynamic limit. 
In this paper we analyse the nature of field and
density fluctuations for the Bose gas with attractive boundary conditions.  
Field fluctuations are centred observables linear in the Bose-field operators 
and density-fluctuations are quadratic in the Bose-fields. 
We derive rigorously the exact form of the 
limiting characteristic functions, \ie we study the limits:
\begin{equation}\label{intro-1}
  \lim_{L \to \infty} \ol{\e^{\ii t F_L}}, \qquad \text{for}\ t \in \R,
\end{equation}
where $L$ indicates the volume dependence of the temperature states $\omega_L$
and the fluctuations $F_L$. 
The thermodynamic limit is computed in various 
ways, leading to the different phases of the Bose gas (normal, critical,
and condensed).  The difference between the critical and condensed phase 
is analysed by a special technique, namely by the interplay between the scaling 
of an external gauge breaking field, used to force the gas into an
extremal state, and the speed at which the chemical 
potential converges. Using this technique, we obtain detailed 
information about the behaviour of the field and density fluctuations 
in the different phases of the Bose gas. 
As turns out, also the fluctuation distributions 
(\ref{intro-1}) and the scaling exponents in the fluctuation observables are 
very sensitive on the boundary conditions and  on the way the thermodynamic 
limit is taken. We prove the existence of different regions in the
space spanned by the scaling parameters of external field and chemical
potential, in which the fluctuations are differently distributed and have a
different degree of abnormality.  The distribution functions also depend 
explicitly on the attractivity parameter of the boundaries. 
The distributions we obtain, are Gaussian or non-Gaussian, normal or abnormal 
depending on different choices of the scaling parameters. 

The main conclusion of this paper can be summarised as follows: details of the
boundary conditions, the strength of the external field, and the thermodynamic
limit have a vast impact not only on the condensation phenomenon, 
but also on the distribution functions of field and density
fluctuations.  These results show that the analysis of the thermodynamic limit should be done 
very carefully and its properties are of utmost importance
for physical observable effects, which are detected at the level of the states
as well as on the fluctuations.  

We remark that we take the thermodynamic limits 
of the states together with the volume scaling of the fluctuation observables.
In this way, the dependence of the limiting distributions and the volume 
scaling exponents on boundary conditions can be made explicit.  These limits are
usually taken separately and boundary conditions are reintroduced with special 
cut-off functions in the definition of the fluctuation operators
\cite{broidioi:1996}.
This type of calculations was previously applied for the classical Curie-Weiss 
Model \cite{verbeure:1994}, but as far as we know not yet for quantum systems.
We obtain explicitly new distribution functions of fluctuations in our Boson
model.  In particular new non-Gaussian critical density fluctuations are 
derived.


\section{The Model}\label{model}
\subsection{The One-dimensional Eigenvalue Problem of the Free Laplacian}
We study the behaviour of a free Bose gas in a $\nu$-dimensional cube
with edges of size $L$, centred around the origin $\Lambda_\nu = [-L/2,
L/2]^\nu$.

The first step in handling this multi-dimensional, many-body system consists of
solving the basic one-dimensional one-body eigenvalue problem of the free 
Laplacian on the Hilbert space $\mathcal{L}^2(\Lambda_1)$: 
\begin{equation}\label{lap1}
-\frac{d^2\psi}{dx^2}(x) = \lambda \psi,
\end{equation} 
with $\psi \in C^2(\Lambda_1)$, the two times continuously differentiable 
complex functions on $\Lambda_1$.
We take the units $\frac{\hbar^2}{2m}=1$ and consider a family of 
self-adjoint extensions of the Laplacian by restricting its domain
using the following boundary conditions 
\begin{align}\label{bound1}
\left [ \frac{d\psi}{dx}(x) - \sigma \psi(x) \right ]_{-\frac{L}{2}} &= 0,\\
\label{bound2}
\left [\frac{d\psi}{dx}(x) + \sigma \psi(x) \right ]_{\frac{L}{2}} &= 0
\end{align}
with $\sigma \in \R$, a parameter governing the elasticity of the
boundaries. Implementing these conditions (\ref{lap1})--(\ref{bound2}), 
gives  the spectrum and eigenfunctions. 
This is worked out in detail in \cite{robinson:1971}, we mention here only 
the results.
It can be proved that there is no
continuous spectrum and that there are infinitely many discrete eigenvalues 
$(\epsilon_n)_{n \in \N}$. The eigenfunctions 
are given by  
\begin{align}\label{even}
\psi_n(x) = \cos(\sqrt{\epsilon_n}x) &\quad 
\text{if $n$ is even}, \\
\label{odd}
\psi_n(x) = \sin(\sqrt{\epsilon_n}x) &\quad 
\text{if $n$ is odd},
\end{align}
up to normalisation.
The spectrum $(\epsilon_n)_n$ depends on the size of the box $\Lambda_1$, and on
the elasticity parameter $\sigma$. 

About the sign of the elasticity parameter,
we distinguish the following situations:
\begin{itemize}
\item {\bf  Neumann boundary conditions ($\sigma = 0$)}
\\
The easiest case, is the case of  Neumann boundary conditions \ie $\sigma =0$.
The system of equations (\ref{lap1})--(\ref{bound2}) can be solved exactly 
and the spectrum is given by
\begin{equation}\label{vns}
\epsilon_n = \left(\frac{n\pi}{L}\right)^2, \qquad \forall n \in \N. 
\end{equation}
The  wave function of the lowest energy level is a constant function.
Physically, this means that the particles in the ground level are not attracted,
nor repulsed by the boundaries, \ie the situation $\sigma = 0$ corresponds to 
perfect elastic boundaries. 

\item {\bf Attractive Boundary Conditions ($\sigma < 0$)}
\\
If $\sigma < 0$, negative eigenvalues are present.  Although an exact solution 
of the set of equations (\ref{lap1})--(\ref{bound2}) is impossible, graphical 
techniques can be used to deduce the following properties of the spectrum.
If $|\sigma | L < 2$ there is only one negative eigenvalue, and the spectrum 
satisfies the following useful spacing properties
\begin{equation}\label{a1s}
\epsilon_0 < - \sigma^2 < 0 < \epsilon_1 < \left(\frac{\pi}{L}\right)^2 
< \epsilon_2 < \left(\frac{2\pi}{L}\right)^2 <
\epsilon_3 < \left(\frac{3\pi}{L}\right)^2 < \epsilon_4 <  \cdots 
\end{equation}
If $|\sigma | L > 2$, the lowest positive eigenvalue becomes negative and there 
are two negative eigenvalues. The spectrum behaves
now as
\begin{equation}\label{a2s}
\epsilon_0 < - \sigma^2 < \epsilon_1 < 0 < \left(\frac{\pi}{L}\right)^2 
< \epsilon_2 < \left(\frac{2\pi}{L}\right)^2 <
\epsilon_3 < \left(\frac{3\pi}{L}\right)^2 < \epsilon_4 <  \cdots 
\end{equation}
Since we are especially interested in the thermodynamic limit with fixed
elasticity parameter $\sigma$, we always assume $L|\sigma | > 2$. The lowest
eigenvalue is monotonically increasing to $-\sigma^2$, and the second eigenvalue
is monotonically decreasing to $-\sigma^2$, and  these processes are 
exponentially fast:
\begin{equation*}
\epsilon_0 = -\sigma^2 - \order{e^{-L|\sigma|}},\quad \epsilon_1 = -\sigma^2 
+ \order{e^{-L|\sigma|}}.
\end{equation*}
It follows from formulas (\ref{even})--(\ref{odd}) that the lowest eigenfunctions
are given by $\cosh$ and $\sinh$ functions. Physically, this means that the
particles in these orbits have a high probability to be close to the 
boundaries, or that the boundaries are attractive.

\item {\bf Repulsive Boundary Conditions ($\sigma > 0$)}  
\\
Using similar techniques as in the case of attractive boundary conditions, it
can be deduced that there are no negative eigenvalues if $\sigma \geq 0$, and 
that the spectrum behaves as
\begin{equation}\label{rs}
0< \epsilon_0 <  \left(\frac{\pi}{L}\right)^2 
< \epsilon_1 < \left(\frac{2\pi}{L}\right)^2 <
\epsilon_2 < \left(\frac{3\pi}{L}\right)^2 < \epsilon_3 <  \cdots 
\end{equation}
The norm of the lowest eigenfunction decreases near the boundaries, or the
boundaries are repulsive.
If $\sigma \uparrow \infty$, we have Dirichlet boundary conditions, the
eigenfunctions vanish at the boundaries and the spectrum can again be solved
exactly, it is given by
\begin{equation}\label{ds}
\epsilon_n = \left(\frac{(n+1)\pi}{L}\right)^2, \qquad \forall n \in \N. 
\end{equation}
\end{itemize}
These considerations about the one-dimensional case can be extended to more 
dimensional systems. 
Let us now consider the $\nu$ dimensional Laplacians on $\Lambda_\nu = 
[-L/2,L/2]^\nu$ with similar domain restrictions, namely, replace
(\ref{bound1})--(\ref{bound2}) with 
\begin{equation*}
\frac{\partial \psi}{\partial n}(x) = \sigma \psi(x), \qquad x \in
\partial\Lambda_\nu,
\end{equation*}
with $\frac{\partial}{\partial n}$ the inward normal derivative.
The eigenvalues are now denoted by $\epsilon_L(k)$,
where the dependence of the eigenvalues on the size of the box is made 
explicit,
$k = (k_1,k_2,\ldots,k_\nu) \in \N^\nu$, and the eigenvalues are given by 
\begin{equation}\label{epsLk}
\epsilon_L(k)= \sum_{i=1}^\nu \epsilon_{k_i}
\end{equation} 
with $\epsilon_{k_i}$ the
$k_i$-est eigenvalue of the one-dimensional free Laplacian on $\Lambda_1$
(\ref{vns})--(\ref{ds}). The eigenfunctions 
$\psi_k(x) \in L^2(\Lambda)$ are products of their one-dimensional components
(\ref{even})--(\ref{odd}).

\subsection{The Bosonic Many Body System }
\subsubsection{The Hamiltonian on the Fock Space}
In order to be able to describe a gas of bosons in
the box $\Lambda = [-L/2,L/2]^\nu$, we use the standard techniques of second quantisation 
\cite{bratteli:1996}
and find the Hamiltonian on the Bose-Fock space 
\begin{equation}\label{HL}
H_L = \sum_{k}(\epsilon_L(k) -\mu_L) a^\dagger(k)a(k) - 
L^{\gamma} h\left(a(0)\mathrm{e}^{-\ii\phi} +
a^\dagger(0)\mathrm{e}^{\ii\phi}\right)
\end{equation}
The index $k$ runs through all vectors in $\N^\nu$, $\epsilon_L(k)$ is the
energy associated with the level $k$ (\ref{epsLk}), $\mu_L \leq \inf_k{\epsilon_L(k)}$ 
is the chemical potential,  
which determines the particle density in the system. The second term
in (\ref{HL}) is a external field term, breaking the gauge symmetry. It is 
added to recover one of the extremal translation invariant equilibrium states 
in the thermodynamic limit. 
This field determines the phase of the condensate, it scales with the volume
with an exponent $\gamma$, which has to be chosen in the range  
\begin{equation}\label{gammarange}
-\nu/2 < \gamma < \nu/2.
\end{equation}
If $\gamma \leq -\nu/2$, its effect is to weak to cause a gauge breaking in the
thermodynamic limit, and if $\gamma \geq \nu/2$, the field is too strong and 
causes an artificial gauge breaking, \ie there is no non-zero critical density.  
$h$ is a positive constant.
The creation operators $a^\dagger(k)$ and their adjoints, the annihilation operators 
$a(k)$ are defined by
\begin{equation}
a(k) = \int\!dx\, \psi_k(x) a(x).
\end{equation}
where $a^\dagger(x)$ is the creation operator of a Boson at $x \in \Lambda_\nu$
 and $a(x)$ is the corresponding annihilation operator. 
The function $\psi_k \in L^2(\Lambda_\nu)$ is the eigenvector by the eigenvalue
$\epsilon_L(k)$ (\ref{epsLk}). 
The local Hamiltonians (\ref{HL}) are diagonalised in terms of quasi-particles,
their creation and annihilation operators are denoted by $b^\dagger$, resp.\ $b$,
and the relation between the annihilation operators of the quasi-particles 
and those of the bare particles is
\begin{equation}\label{ba}
\left \{ 
   \begin{array}{lcl} 
     b(0) & = & a(0) - \frac{L^\gamma h \mathrm{e}^{i\phi}}{\epsilon_L(0) -\mu_L}, \\
     b(k) & = & a(k), \qquad  \forall k \in \N^\nu \backslash \{0\}. 
   \end{array} 
\right .
\end{equation} 
This enables to rewrite the Hamiltonians (\ref{HL})  as
\begin{equation}\label{HLD}
H_L = \sum_{k}(\epsilon_L(k) -\mu_L) b^\dagger(k)b(k) -
\frac{L^{2\gamma}h^2}{\epsilon_L(0) -\mu_L}.
\end{equation}
We see that in terms of the quasi-particles the Hamiltonians $H_L$ are
essentially the free Hamiltonians plus an unimportant constant
term. The volume dependence of the spectrum and the chemical potential are
important in the remainder of this paper.

\subsubsection{Quasi Free Equilibrium States}
The finite volume equilibrium states or Gibbs states $\omega_L$ are now 
easily established using the expression in terms of the quasi-particles 
(\ref{HLD}). The equilibrium states then coincide
with the expression for the equilibrium states of the free Bose gas without 
external field \cite{bratteli:1996}. 
These states are quasi-free states, the easiest way to characterise them is 
to use the truncated correlation functions $\omega_L(\cdots)_T$, recursively 
defined, using the following expression,
for all $A_i,\ i =1,2,\ldots$ creation or annihilation operators  or combinations of them:
\begin{equation}\label{trf}
\omega_L(A_1\cdots A_n) = \sum_{\tau \in \mathcal{P}_n} \prod_{J \in
\tau}\omega_L(A_{j(1)},\cdots,A_{j(|J|)})_T,
\qquad \forall n \in \N,
\end{equation}
where the sum $\tau \in \mathcal{P}_n$ runs over all ordered partitions $\tau$ of 
a set of $n$ elements in subsets $J = \{j(1),\ldots, j(|J|)\} \in \tau$.
The truncated functions associated with the equilibrium states $\omega_L$ 
satisfy
\begin{equation}\label{oL}
\begin{split}
&\omega_L(b^{\sharp}(f))_T= 0 \\
&\omega_L(b^\dagger(f_1),b^\dagger(f_2))_T = \omega_L(b(f_1),b(f_2))_T = 0 \\
&\omega_L(b^\dagger(f_1),b(f_2))_T = \sum_{k}
\overline{\hat{f_2}(k)}\hat{f_1}(k)\frac{1}{e^{\beta(\epsilon_L(k)-\mu_L)}-1}\\
 &\omega_L(b^{\sharp}(f_1),\cdots,b^{\sharp}(f_n))_T = 0 \qquad n \geq 3.
\end{split}
\end{equation}
with $f_1,f_2,\ldots \in L^2(\Lambda_\nu)$, and $b^\sharp$ can be either $b$ or
$b^\dagger$. Only the two-point functions are non-zero,
the Fourier transform $f(x)
\mapsto \hat{f}(k)$, $k \in \N^\nu$ is defined by
\begin{equation}
\hat{f}(k) = \langle \psi_k | f \rangle =  
\int \! dx \ \overline{\psi_k(x)}f(x).
\end{equation}

\section{Condensation}\label{condensation}
In this section we prove the existence of different phases in the
thermodynamic limit of the system, \ie we prove the existence of a
Bose-Einstein condensate -- a macroscopic occupation of the lowest energy level
-- when the chemical potential $\mu_L$ scales correctly.    
Here we take the thermodynamic limit $(L \uparrow \infty)$ with
fixed chemical potentials $\mu_L$ and varying densities 
(as in \cite{robinson:1976}), rather than 
with a fixed density and varying chemical potential (as in \cite{landau:1979,
bratteli:1996}).  
But before discussing this in detail, let us first prove 
two lemmata.

\begin{lemma}\label{lemma1}
Choosing the chemical potentials $\mu_L$ for each volume $\Lambda_\nu =
[-L/2,L/2]^\nu$ 
equal to 
\begin{equation}\label{muL}
\mu_L = \epsilon_L(0) - \frac{h}{\sqrt{\rho_0}}L^{-\alpha_*}
\end{equation} 
where the scaling exponent $0 < \alpha_* = \nu/2 - \gamma$ and $\rho_0 \in \R^+$, 
yield a non-zero
occupation of the lowest energy level in the thermodynamic limit, \ie
with this choice for the series $(\mu_L)_L$ we have
\begin{equation}\label{rho0}
\lim_{L \to \infty} L^{-\nu}\omega_L(a^\dagger(0) a(0) ) = \rho_0
\end{equation}
and the gauge-invariance of the limiting state is broken, \ie
\begin{equation}\label{gbreak}
\lim_L L^{-\nu/2}\omega_L(a(0)) = \sqrt{\rho_0}\mathrm{e}^{\ii\phi}.
\end{equation}
\end{lemma}
{\bf Proof}\\
The density of particles in the lowest energy level in the equilibrium state 
$\omega_L$ for a finite volume $\Lambda_\nu $ (\ref{oL}) is given by
\begin{equation*}
\rho_0(L) = L^{-\nu}\omega_L(a^\dagger(0) a(0)).
\end{equation*}
This expression can be written in terms of the quasi-particles using the 
relations (\ref{ba}) and the expression for $\mu_L$ (\ref{muL}), \ie use
$a(0) = b(0) + L^{\nu/2}\sqrt{\rho_0} \mathrm{e}^{\ii\phi}$ to find that
\begin{equation*}
\rho_0(L) = L^{-\nu}\omega_L(b^\dagger(0)b(0))
+ L^{-\nu/2} \sqrt{\rho_0}\left(\omega_L(a(0))\mathrm{e}^{-\ii\phi} + 
\omega_L(a^\dagger(0))\mathrm{e}^{\ii\phi}\right) - \rho_0.
\end{equation*}
The first term is of order $\order{L^{-\nu/2 - \gamma}}$, this can be seen using
the two-point functions (\ref{oL}) as follows 
\begin{eqnarray*}
L^{-\nu}\omega_L(b^\dagger(0) b(0)) & = & L^{-\nu} \frac{\e^{ - \beta( 
\epsilon(0) - \mu_L)}}{1 - \exp^{ - \beta( \epsilon(0) - \mu_L)}}\\
& = &  L^{-\nu} \frac{ 1  - \order{L^{-\alpha_*}}}{1 - 1 + \frac{\beta
h}{\sqrt{\rho_0}} L^{-\alpha_*} + \order{L^{-2\alpha_*}}}\\
& = & \order{L^{\alpha_* - \nu}} = \order{L^{-\nu/2 - \gamma}}.
\end{eqnarray*}
And by (\ref{gammarange}), this term vanishes in the thermodynamic limit. 
We use this observation to establish a bound on the difference between $\rho_0(L)$ and
$\rho_0$
\begin{equation*}
0 \leq \rho_0(L) - L^{-\nu/2} \sqrt{\rho_0}(\omega_L(a(0))e^{-\ii\phi} +
\omega_L(a^\dagger(0))e^{\ii\phi})
+  \rho_0  \leq \order{L^{-\nu/2 - \gamma}}.
\end{equation*}
Using the Schwarz inequality $L^{-\nu/2}|\omega_L(a(0))| \leq 
\sqrt{L^{-\nu}\omega_L(a^\dagger(0)a(0))} = \sqrt{\rho_0(L)}$
the above expression is transformed into
\begin{eqnarray*}
\rho_0(L) - 2\sqrt{\rho_0}\sqrt{\rho_0(L)} + \rho_0 & \leq & 
\order{L^{-\nu/2 - \gamma}}\\
(\sqrt{\rho_0} - \sqrt{\rho_0(L)})^2 & \leq & 
\order{L^{-\nu/2 - \gamma}}
\end{eqnarray*}
and hence,
\begin{eqnarray*}
|\sqrt{\rho_0} - \sqrt{\rho_0(L)}| & \leq & 
\order{L^{-\nu/4 - \gamma/2}}\\
|(\sqrt{\rho_0} - \sqrt{\rho_0(L)})(\sqrt{\rho_0} + \sqrt{\rho_0(L)})| 
& \leq & \order{L^{-\nu/4 - \gamma/2}}\\
|\rho_0 - \rho_0(L)| & \leq & \order{L^{-\nu/4 - \gamma/2}}.
\end{eqnarray*}
Indicating that $\rho_0(L)$ converges to $\rho_0$ in the thermodynamic limit.

Also the second part of the lemma (\ref{gbreak}) can be proved by means of the 
Schwarz inequality, 
\begin{equation*}
L^{-\nu}|\omega_L(b(0))|^2 \leq L^{-\nu}\omega_L(b^\dagger(0)b(0)) 
= \order{L^{-\nu/2 -\gamma}}. 
\end{equation*} 
Using the relations (\ref{ba}) and the expression for the chemical potential, 
this yields
\begin{equation}
\lim_{L \to \infty} L^{-\nu/2}\omega_L(a(0)) = \sqrt{\rho_0}e^{\ii\phi}.
\end{equation}
\begin{flushright}
$\square$
\end{flushright}
The result (\ref{rho0}) is not yet a complete proof of Bose-Einstein condensation,
it should also be proved that the total particle density is finite with this 
choice for the chemical potential. Only then one can speak of a macroscopic 
fraction of particles condensed in the lowest energy level.

The total density $\rho_\nu(L)$ in a volume $\Lambda_\nu$  
in the equilibrium state $\omega_L$ is easily derived from the relations
(\ref{ba}) and the two-point functions (\ref{oL})
\begin{equation}\label{density}
\rho_\nu(L) = L^{-\nu}\sum_k \omega_L(a^\dagger(k)a(k)) = \rho_0(L) +
L^{-\nu}\sum_{k\ne 0} \frac{z_L e^{-\beta \epsilon_L(k)}}{1-z_L e^{-\beta
\epsilon_L(k)}}
\end{equation}
where $z_L$ is the activity or fugacity $z_L = \mathrm{e}^{\beta \mu_L}$, and 
the sum over $k$ runs over all $k \in \N^\nu\backslash\{0\}$.
We formulate now the well-know result
\begin{lemma}\label{lemma2}
Take $z_L \to z \leq \lim_L e^{\beta\epsilon_L(0)}$, then
\begin{equation}
\lim_{L \to \infty} L^{-\nu}\sum_{k\ne 0} \frac{z_L e^{-\beta \epsilon_L(k)}}
{1-z_L e^{-\beta \epsilon_L(k)}} = \lambda^{-\nu} J(\nu/2,z)
\end{equation}
where $\lambda = \pi \sqrt{\beta}$ is the thermal wavelength. 
The function 
\begin{equation*}
J(\nu/2,z) = \int_{x_i >0}\!\!\! \mathrm{d}^\nu x \frac{z\mathrm{e}^{-x^2}}
{1-z\mathrm{e}^{-x^2}}
\end{equation*}
is the Jonqui\`ere function. This result is valid for any type of boundary 
conditions specified in (\ref{bound1})--(\ref{bound2}).
\end{lemma}
{\bf Proof}\\
Based on the spacing properties of the eigenvalues (\ref{vns})--(\ref{ds}) and 
the convergence of Riemann sums to Riemann integrals. A proof can be found in 
\cite{robinson:1976}. 
\begin{flushright}
$\square$
\end{flushright}
The Jonqui\`ere function $J(\nu/2,z)$ can be expressed as 
\begin{equation}
J(\nu/2,z) = \sum^\infty_{n=1}\frac{z^n}{n^{\nu/2}}
\end{equation}
It is an analytic function of $z$ in the cut-plane, the cut being from $1$ to
$\infty$ along the positive real $z$-axis. In the limiting case $z \to 1$, it
converges to the Riemann zeta function $\zeta(\nu/2)$, this is finite for
dimensions higher than three or $\nu \geq 3$ \cite{magnus:1966}. 

Combining the results of the two lemmata we can analyse the different phases of
the Bose gas. If we take the thermodynamic limit such that the series 
$(\mu_L)_L$ converges to a certain value $\mu$, strictly lower than the limit 
of the lowest eigenvalue $ \mu < \mu_* = \lim_{L \to \infty} \epsilon_L(0)$, 
there is no macroscopic occupation of the lowest energy level (cfr.\ lemma \ref{lemma1}),
and the total density converges to a finite value (lemma \ref{lemma2}). 
This situation is called the normal phase.

The critical and condensed phase can be reached if we take a 
series $(\mu_L)_L$ converging to the maximal value $\mu_*$, provided that 
the total density (\ref{density}) is finite in this limit, \ie in dimensions 
$\nu \geq 3$ if we have $\sigma \geq 0$ boundary conditions (where $\mu_* = 0$), and in al
dimensions for $\sigma < 0$ boundary conditions (where $\mu_* = -\nu\sigma^2 <
0$).  As said before, we study in this paper the  critical and condensed phase
in the thermodynamic limit ($L\uparrow \infty$) taking series $(\mu_L)_L$  of the form
\begin{equation}\label{mul-series}
\mu_L = \epsilon_L(0) - c L^{-\alpha},
\end{equation}
where $c > 0$ some positive constant and $ \alpha > 0$ a scaling exponent.
Clearly, these series (\ref{mul-series}) converge to $\mu_* =
\lim_L\epsilon_L(0)$.
Depending on the scaling exponent $\alpha$, we have a different phase in the
thermodynamic limit.

If the convergence is slow, \ie if the scaling exponent $\alpha$ is 
chosen in the range 
\begin{equation}\label{ac}
0 < \alpha < \alpha_* = \nu/2 -\gamma,
\end{equation}
we end up in the critical phase. The total density converges to its
critical value $\lambda^{-\nu}J(\nu/2,z_*)$, with $z_* = \mathrm{e}^{\beta\mu_*}$. 
There is no macroscopic occupation of the lowest level, since the convergence 
$\mu_L \to \mu_*$ is too slow (lemma \ref{lemma1}).

If the convergence is sufficiently fast, \ie if
the scaling exponent $\alpha$ (\ref{mul-series}) is large enough, 
\begin{equation}\label{a}
\alpha = \alpha_* = \nu/2 -\gamma,
\end{equation}
\ie if $\mu_L$ converges to $\mu_*$ as specified in lemma \ref{lemma1},  
the total density can reach arbitrary values above 
the critical density $\lambda^{-\nu}J(\nu/2,z_*)$.  There is a non-zero density
of the condensate $\rho_0$. 
This condensate density is determined by the constants $c$ (\ref{mul-series}) 
and $h$ of the external field (\ref{HL}), \ie $\rho_0 = h^2/c^2$, 
cfr.\ equation (\ref{muL}).

The situation with $\mu_L = \epsilon_L - cL^{-\alpha}$
with $\alpha > \nu/2 - \gamma$, is not physically meaningful since we end up
in a situation where the density of the condensate $\rho_0(L)$ 
diverges in the thermodynamic limit. 

Let us summarise this in the following.

\begin{theorem}[Phases of the Free Bose Gas]
Fix a temperature $\beta$, the thermodynamic limit of the $\beta$-equilibrium 
states $\omega_L$, \ie the limit $(L \uparrow
\infty)$ with $(\mu_L \to \mu)$, exists if the total density (\ref{density}) 
converges to a finite value $\rho$, and 
\begin{itemize}
\item $\rho  < \lambda^{-\nu}J(\nu/2,z_*)$, if $\mu_L \to \mu < \mu_* = \lim_L \epsilon_L(0)$
(Normal Phase),
\item $\rho =  \lambda^{-\nu}J(\nu/2,z_*)$, if $\mu_L = \epsilon_L(0) 
- cL^{-\alpha}$ with $c > 0$ and
$
 0 < \alpha < \nu/2 - \gamma, 
$ 
and if $J(\nu/2,z)$ converges (Critical Phase), 
\item $\rho = \rho_0 + \lambda^{-\nu}J(\nu/2,z_*)$, if $\mu_L = \epsilon_L(0) -
\frac{h}{\sqrt{\rho_0}}L^{-\alpha_*}$
with 
$
 \alpha_* = \nu/2 - \gamma, 
$
provided that $J(\nu/2,z)$ converges (Condensed Phase).
\end{itemize}
\end{theorem}
For attractive boundary conditions, the condensation phenomenon takes place in
any dimension, but due to the special form of the wavefunctions of the
lowest energy levels, the condensate is localised near the boundaries, and the
condensation is a pure surface effect. For $\sigma \geq 0$, the condensation is a
bulk phenomenon and only takes place in dimensions higher than three 
$(\nu \geq 3)$.  
More detailed discussions of these different types of condensation can be found
in \cite{robinson:1976,landau:1979,bratteli:1996}.

It is clear that the strength parameter of the external field $\gamma$ plays an
important role in the condensed and critical phases. If $\gamma$ is small, the
$\alpha$ has to be larger, or $\mu_L$ has to converge faster to zero, in order
to have condensation; for larger values of $\gamma$, condensation is already
present at a slower convergence rate for $\mu_L$. A value $\gamma = \nu/2$
provokes an artificial gauge breaking, and there is always condensation at any
density. If $\gamma \leq -\nu/2$, the nature of the phase transition changes,
the condensate is no longer of a well defined phase, and we do not find a 
single extremal state in the thermodynamic limit, but a mixture. Hence
clustering is absent and we can no longer analyse the fluctuations. The
interplay between $\gamma$ and $\alpha$ plays also its role below
where we analyse the field and density fluctuations.  


\section{Field Fluctuations}
In this section we study the scaling behaviour of field fluctuations, or
fluctuations of operators of the form
\begin{eqnarray}
A_k^+ & = & \frac{1}{\sqrt{2}}\left( a(k) + a^\dagger(k) \right); \\
A_k^- & = & \frac{\ii}{\sqrt{2}}\left( a(k) - a^\dagger(k) \right), 
\end{eqnarray}
with $k \in \N^\nu$, different modes.
The local k-mode field fluctuations are defined by
\begin{equation}
F_{L,\delta}(A^\pm_k) =L^{-\delta} \left ( A^\pm_k - \omega_L(A^\pm_k) \right) 
\end{equation}
$\delta$ is the scaling exponent and should be chosen such that the limiting
characteristic function 
\begin{equation}\label{char}
\varphi(F_\delta(A^\pm_E)) :  t \mapsto
\varphi(F_\delta(A^\pm_E))(t) \equiv \lim_{L\to \infty} 
\omega_L(\exp(\ii tF_{L,\delta}(A^\pm_{k_L}))),
\end{equation}
is non-trivial \cite{broidioi:1996}.
The limits we consider here are limits of series of local fluctuations
$F_{L,\delta}(A^\pm_{k_L})$, where the vectors $k_L \in \N^\nu$ are chosen
such that we can associate with the series $\left(k_L\right)_L$ a series of eigen\-values
$\left(\epsilon_L(k_L)\right)_L$, converging to a certain value $E \in
\{-\nu\sigma^2\} \cup  [-(\nu -1)\sigma^2,\infty)$ in the limit $L \uparrow
\infty$ . The scaling exponent $\delta$ 
depends on $E$ and on the phase of the Bose gas.
 
In the case of field fluctuations, the limiting characteristic function can 
explicitly be obtained, due to the quasi-free character of the states
$\omega_L$.
\begin{lemma}
Field fluctuations are Gaussian
\begin{equation}\label{ffg}
\varphi(F_\delta(A^\pm_E))(t) = \lim_{L \to \infty} 
\exp\left(-\frac{t^2}{4}L^{-2\delta}
\left(1+ 2\omega_L(b^\dagger(k_L)b(k_L))\right)\right).
\end{equation}
\end{lemma} 
{\bf Proof}\\
The characteristic functions (\ref{char}) can be written in the following
expansion \cite{bratteli:1996}
\begin{equation}\label{charexp}
\omega(\exp(\ii t Q)) = \exp \sum_{n=1}^\infty
\frac{(\ii t)^n}{n!}\omega_L(\underbrace{Q,Q,\ldots,Q}_{n\ \text{times}})_T, 
\end{equation}
with $\omega_{L}(Q,Q,\ldots,Q)_T$, the $n$-point truncated correlation functions 
(\ref{trf}).
In the case of field fluctuations, \ie if we substitute $F_{L,\delta}
(A^\pm_k)$ for $Q$ in (\ref{charexp}), due to the properties of the equilibrium
states $\omega_L$ (\ref{oL}), only the 
second order term in this expansion is different from zero.
This term can be rewritten in terms of the quasi-particles as
\begin{eqnarray*}
\omega_{L}(F_{L,\delta}(A^\pm_k),F_{L,\delta}(A^\pm_k))_T &=& \omega_L(F_{L,\delta}
(A^\pm_k)^2) - \omega_L(F_{L,\delta}(A^\pm_k))^2\\
&=& \frac{L^{-2\delta}}{2}\left(1+ 2\omega_L(b^\dagger(k_L)b(k_L)) \right)  
\end{eqnarray*}
and this yield the explicit form of the characteristic function (\ref{ffg}).
\begin{flushright}
$\square$
\end{flushright}
In spite of the fact that field fluctuations are Gaussian, they can be abnormal,
in the sense that there has to be a non-zero scaling exponent $\delta$, in order
to have non-trivial fluctuations.

\begin{theorem}[Field Fluctuations]\label{fifl}
The limiting characteristic functions of the k-mode field fluctuations
\begin{equation}
\varphi(F(A^\pm_E)(t) = \lim_{L \to \infty}
\omega_L\left(\exp(\ii tF_{L,\delta}(A^\pm_{k_L}))\right) 
\end{equation}
with $\epsilon_L(k_L) \to E$,
tend to non-trivial distributions if 
\begin{itemize}
\item $\delta = 0$ in the normal phase $(\mu_L \to \mu < \mu_* = -\nu\sigma^2)$ 
\begin{equation}\label{cf1}
\varphi(F(A^\pm_E))(t) = \exp(-\frac{t^2}{4} \coth(\beta/2(E-\mu)))
\end{equation}
\item $\delta = 0$ in the critical and condensed phase if 
$E \ne -\nu\sigma^2$
\begin{equation}\label{cf2}
\varphi(F(A^\pm_E))(t) = \exp(-\frac{t^2}{4} \coth(\beta/2(E+ \nu\sigma^2))
\end{equation}
\item $\delta = \alpha/2$ in the critical and condensed phase $(\mu_L =
\epsilon_L(0) - cL^{-\alpha})$, for all $\alpha$: $0 < \alpha \leq \nu/2 -\gamma$ 
and if $\epsilon_L(k_L) \to E = -\nu\sigma^2$
\begin{equation}\label{cf3}
 \varphi(F(A^\pm_{-\nu\sigma^2}))(t) = \exp(-\frac{t^2}{2} \frac{1}{\beta c}).
\end{equation}
In the condensed phase, \ie if $\alpha  = \nu/2 -\gamma$
this expression can also be expressed in function of the condensation density
$\rho_0$, \ie
\begin{equation}\label{cf4}
\varphi(F(A^\pm_{-\nu\sigma^2}))(t) = \exp(-\frac{t^2}{2}
\frac{\sqrt{\rho_0}}{\beta h}).
\end{equation}
\end{itemize}
\end{theorem}
{\bf Proof}\\
The two-point function appearing in the expression 
for the distribution of field fluctuations (\ref{ffg}) is given by
\begin{equation}\label{bkbk}
\omega_L(b^\dagger(k_L)b(k_L)) = \frac{\exp(-\beta(\epsilon_L(k_L)-\mu_L))}
{1 - \exp(-\beta(\epsilon_L(k_L)-\mu_L))}. 
\end{equation}
This function converges in the limit $(L \uparrow \infty)$ if 
\begin{equation*}
\lim_{L \to \infty} \left| \epsilon_L(k_L)-\mu_L \right| > 0. 
\end{equation*}
 This situation occurs in the normal phase for all $\epsilon_L(k_L) \to E$, 
 but in the critical and condensed phase only for $\epsilon_L(k_L) \to E \ne
 -\nu\sigma^2$. It yields normal fluctuations, \ie    
$\delta =0$, and the explicit form of the distributions (\ref{cf1}) and 
(\ref{cf2}).

Let us now consider the case $E = -\nu\sigma^2$ in the critical or condensed
phase. 
If $E =  -\nu\sigma^2$, the vectors $k_L$ should be in
$\{0,1\}^\nu$ for sufficiently large $L$, only in that case 
there is convergence $\epsilon_L(k_L) \to -\nu\sigma^2$, and this convergence 
is exponentially fast. In the critical phase $\mu_L$ also
converges to $-\nu\sigma^2$ (\ref{mul-series}), and the expectation value
appearing in expression (\ref{bkbk})  diverges.
An extra scaling $L^{-\delta}$ is necessary in order to obtain 
finite variances.  
The highest order term  of (\ref{bkbk}) behaves as
\begin{eqnarray*}
2L^{-2\delta}\omega_L(b^\dagger(k_L)b(k_L)) & = & 2L^{-2\delta}
\frac{\e^{-\beta(\epsilon_L(k_L)-\mu_L)}}
{1 - \e^{-\beta(\epsilon_L(k_L)-\mu_L)}} \\
& = &  2L^{-2\delta}\frac{1  - \beta cL^{-\alpha} + \order{L^{-2\alpha}} 
+ \ldots}{1 - 1 + \beta cL^{-\alpha}  
+ \order{L^{-2\alpha}} 
+ \ldots}\\
& = & 2L^{-2\delta} \Big( \frac{1}{\beta c} L^{\alpha} + \order{1} + \ldots
\Big).
\end{eqnarray*}
Hence, a scaling $\delta = \alpha/2$ is needed in order to avoid divergences or 
trivial distributions. The distribution then converges to
 \begin{equation*}
 \varphi(F(A^\pm_{-\nu\sigma^2}))(t) = \exp\Big(-\frac{t^2}{2} \frac{1}{\beta
 c}\Big).
\end{equation*}
In the condensed phase, \ie if $\alpha = \nu/2 -\gamma$, we can replace 
$c$ by an expression depending on the condensate density and the external
field $h$: $ \frac{1}{\beta c} = \frac{\sqrt{\rho_0}}{\beta h}$ (\ref{cf4}). 
\begin{flushright}
$\square$
\end{flushright}


\section{Density Fluctuations}

The local density fluctuations $F_{L,\delta}(N_0)$ are defined by
\begin{equation}\label{FN0}
F_{L,\delta}(N_0) = L^{-\nu/2 -\delta}
\int_\Lambda \! \mathrm{d}x\ a^\dagger(x)a(x)
-\omega_L(a^\dagger(x)a(x)), 
\end{equation}
or equivalently
\begin{equation}
F_{L,\delta}(N_0) = L^{-\nu/2 -\delta}
 \sum_{p \in \N^\nu} a^\dagger(p)a(p)
-\omega_L(a^\dagger(p)a(p)).
\end{equation}
it turns out these density fluctuations can be divergent in the critical and
condensed phases, and suitable non-zero scaling exponents $\delta$ have to 
be added. 
 
Instead of using such an extra scaling factor, one can introduce 
modulated fluctuations. 
Mostly one chooses a cosine function as modulation function, and one introduces
the so called $k$-mode fluctuations \cite{verbeure:1994,michoel:1999b} by:
\begin{equation*}
F_k(A) = L^{-\nu/2} \int_\Lambda \! \mathrm{d}x\ ( \tau_x(A)-\ol{\tau_x(A)})\cos(kx).
\end{equation*}
Inspired by \cite{michoel:1999b}, where the Goldstone phenomenon was 
studied in interacting Bose gases, we consider here $k$-mode density 
fluctuations given by 
\begin{multline}\label{FNK}
F_{L,\delta}(N_k) =L^{-\nu/2-\delta} \frac{1}{2}\sum_{p} \Big(  a^\dagger(p)a(p+k) +
a^\dagger(p+k)a(p)
\\
-\ \omega_L(a^\dagger(p)a(p+k) + a^\dagger(p+k)a(p)) \Big).
\end{multline}
These observables can be considered as exchange correlations between particles 
of momentum differences $k$, but also as modulated density fluctuations.
In the case of periodic boundary conditions, which was considered in
\cite{michoel:1999b}, this form of modulation coincides with the $k$-mode 
cosine-modulated fluctuations. Here, in the case of elastic boundary conditions,
these fluctuations (\ref{FNK}) correspond to a more complicated form of
modulation
\begin{equation*}
F_{L,\delta}(N_k)=L^{-\nu/2-\delta} \int_\Lambda \! \mathrm{d}x \! \int_\Lambda \! 
\mathrm{d}y\ a^\dagger(x)a(y) \bigg(\sum_p
\overline{\psi_p(x)}\psi_{p+k}(y)/2\bigg) + h.c.
\end{equation*}
The modulation function $\sum_p\cdots$ is here no longer of a delta-function
type, \ie  $\delta(x-y)\cos(kx)$, but it is a more smeared out function.  

These fluctuations (\ref{FN0})--(\ref{FNK}) have no longer a priori a Gaussian
distribution. Nevertheless, we are still able to compute rigorously and
explicitly their distribution functions as a function of temperature, density,
external field strength and boundary conditions.

For the sake of compactness of the paper, we will not discuss in this section 
the case where $k$ depends on the volume $k = k_L$ (as in Theorem \ref{fifl}), 
but take $k$ constant. The more general situation however is easily obtained 
from this case.

\subsection{Quasi-particles density fluctuations}
The distributions of the density fluctuations of the bare particles is related
to the distributions of similar density fluctuations in terms of the
quasi-particles (\ref{ba}). The first step in calculating the distribution of
density fluctuations consists in
finding the distributions of the density fluctuations of the quasi-particles,
\begin{multline}\label{FNK'}
F_{L,\delta}(N_k') =L^{-\nu/2-\delta} \frac{1}{2}\sum_{p} \Big(  b^\dagger(p)b(p+k) 
+ b^\dagger(p+k)b(p) 
\\ -\ \omega_L(b^\dagger(p)b(p+k) + b^\dagger(p+k)b(p)) \Big),
\end{multline}
where the accent in $N_k'$ is added to make a distinction with the
particle density of the bare particles.  


\subsubsection{$k \ne 0$ quasi-particle density fluctuation}

First we consider the $k \ne 0$ density fluctuations.
In this case, the expectation value appearing in (\ref{FNK'}) is zero and may be
left out.
If we want to calculate the characteristic function 
$$\varphi(t): t \mapsto \lim_{L\to \infty} \ol{\e^{\ii t F_{L,\delta}(N_k')}} $$ 
we can use expansion (\ref{charexp}) in the $n$-point truncated 
correlation functions:  
\begin{equation*}                  
\ol{F(N_k'),\ldots,F(N_k')}_T. 
\end{equation*} 

These $n$-point truncated functions can be calculated, but first we need to know 
something about the expectation values of the powers of $F(N_k)$. 

\begin{figure}[h] 
\begin{center}
\includegraphics{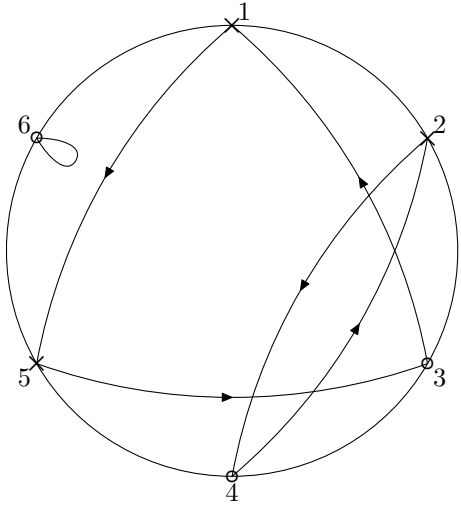}
\end{center}
\caption{A possible configuration with $6$ sites and $3$ cycles.}
\label{circ1}
\end{figure}

\paragraph{The expectation value of powers of the quasi particle density 
fluctuations.}
Consider the function $\ol{F(N_k')^{n}}$
with $F(N_k')$ as in (\ref{FNK'}).
This function can be expanded in $2^{n}$ functions,  
if every factor $F(N_k')$ is splitted into two parts
\begin{equation}
\left(L^{-\nu/2}/2\sum_p b^\dagger(p)b(p+k)\right) + 
        \left(L^{-\nu/2}/2\sum_p b^\dagger(p+k)b(p)\right).
\end{equation}
Those $2^n$ terms are all of the form
\begin{equation}\label{1-term}
L^{-n\nu/2 -n\delta}\frac{1}{2^n}\sum_{p_1,\ldots, p_n} \ol{b^\dagger(p_1)b(p_1+k)\cdots 
b^\dagger(p_n + k)b(p_n)}
\end{equation}
where $n$ factors appear being either $b^\dagger(p_i)b(p_i+k)$ or 
$b^\dagger(p_i + k)b(p_i)$, $\forall i =1,2,\ldots,n$.
Such a term (\ref{1-term}) can be represented as a configuration of
$n$ symbols on a circle, e.g. see (Fig.~\ref{circ1}) with $n = 6$.

We draw a circle and add $n$ points, choose a starting point and count
the different sites clockwise. To every site $1,2,3,\ldots,n$ we add a symbol 
being either $\circ$ or $\times$. We draw a '$\circ$' at site $i$, for a factor
$b^\dagger(p_i)b(p_i+k)$, and a '$\times$' for a factor $b^\dagger(p_i +
k)b(p_i)$.
In the example (Fig.\ \ref{circ1}), we represent the monome
\begin{multline}\label{expr1}	
\sum_{p_1,\ldots,p_6}
\omega_L \left(
b^\dagger(p_1 +k)b(p_1)b^\dagger(p_2 +k)b(p_2)b^\dagger(p_3)b(p_3+k)
\right. \\ \left.
b^\dagger(p_4)b(p_4+k)b^\dagger(p_5 +k)b(p_5)b^\dagger(p_6)b(p_6+k)
\right).
\end{multline}
Using the quasi-freeness of the states (\ref{oL}), these functions (\ref{1-term})
can be expressed as a sum of terms consisting of $n$ products
of 2-point functions only. All partitions into ordered pairs appear once in
this expansion.
A possible term  in this expansion of products of two-point functions is now
represented by a directed graph on this circle, connecting the different sites with each
other, obeying the rule that on every site there must start an arrow and must
end one.  Every arrow corresponds to a two-point function. Such a two-point
function is constructed by combining
the creation operator from the site where the arrow starts with the annihilation
operator from the end-point. The order of the operators is defined by the order
of the sites $1,2,\ldots,n$. A loop is also possible,  then we combine the
creation operator with the annihilation operator from the same site, \eg in the
above example (Fig.~\ref{circ1}), there is a loop at site $6$ , this yields a factor
$\ol{b^\dagger(p_6)b(p_6+k)}$. Since $k \ne 0$, such loop-factors are zero by
gauge-invariance of the states $\omega_L$ (\ref{oL}). 
A graph on a circle consists of separated connected graphs on subsets 
or cycles, \eg the  graph in (Fig.~\ref{circ1}) consists of a two-point cycle,
a three-point cycle and a one-point cycle or loop. 
The total term represented in (Fig.~\ref{circ1}) reads
\begin{multline}\label{expr2}
  \sum_{p_1,\ldots,p_6} \ol{b^\dagger(p_1 +k)b(p_5)}\ol{b(p_3 +k
  )b^\dagger(p_5+k)}
  \ol{b(p_1)b^\dagger(p_3)} 
 \\
  \ol{b^\dagger(p_2+k)b(p_4+k)}\ol{b(p_2)b^\dagger(p_4)}
  \ol{b^\dagger(p_6)b(p_6+k)}.
\end{multline}
A number of useful properties can immediately be deduced from this
representation:

\begin{itemize}

\item Every (non-zero) cycle has only one summation index. 

Every two-point function yields a relation between the summation indices of the
creation and annihilation operator.  All types of two point functions appearing
in these expressions (\ref{expr2}) are zero unless the indices of the creation
and the annihilation operators are the same. This yields a linear relation
between the summation indices $p_1,p_2,\ldots$ in the cycle, and hence only 
one summation is free.

\item A cycle containing not the same numbers `$\circ$' as `$\times$' are zero,
consequently cycles over an odd number of sites are always zero.

The sum of the indices of the creation operators minus the sum of the indices 
of the annihilation operators in a cycle must be zero, otherwise there
will always be a two-point function where the indices of the two operators 
are different, and such a factor is zero. In cycles where there are not as many
symbols `$\circ$' as `$\times$', this sum is always different from zero.
\end{itemize}


\paragraph{ Truncated functions.}
The following step in the calculation of the characteristic function, consists 
in the calculation of the $n$-point truncated correlation function itself.


\begin{lemma}\label{odd-tf}
All odd truncated functions vanish.
\begin{equation*}
\omega_L \big( \underbrace{F(N_k'),\ldots,F(N_k')}_{2n+1\ factors}\big)_T = 0, \qquad \forall n
\in \N.
\end{equation*} 
\end{lemma}
{\bf Proof}

First note that the one-point function vanishes,
$$
\omega_L(F(N_k'))_T = L^{-\nu/2}/2\sum_p\ol{b^\dagger(p)b(p+k)} 
+ \ol{b^\dagger(p+k)b(p)}= 0
$$
due to the gauge invariance of the states $\omega_L$. 

Consider now the $2n + 1$-point truncated function, and suppose that all $m$-point 
truncated functions where $m$ is an odd number less than $2n + 1$, vanish.  
Using the definition of the truncated functions (\ref{trf}), the $2n + 1$-point
truncated function is written as
\begin{equation}
\ol{F(N_k'),F(N_k'),\ldots,F(N_k')}_T = \ol{{F(N_k')}^{2n +1}} - 
\sum_{\tau \in \mathcal{P'}}\prod_{J \in \tau} \ol{F(N_k'),\ldots,F(N_k')}_T,
\end{equation}
where the sum in the second term on the rhs runs through all partitions 
$\tau \in \mathcal{P'}$ in
two or more subsets $J$ of a string of $2n + 1$ elements. 
Each term in this sum on the rhs contains at least one factor with a truncated
function over an odd number of points $m < 2n + 1$, and such a factor is zero 
by induction hypothesis.

Also the first term $\omega_L({F(N_k')}^{2n +1})$ vanishes. 
Expand this term using the definition of  $F(N_k')$ (\ref{FNK'}),  
as in (\ref{1-term}). All possible configurations in terms of two-point
functions (Fig.\ \ref{circ1}), contain at least one cycle with not as many 
symbols `$\circ$' or `$\times$', because the total number of symbols is odd, 
and such cycles vanish. Hence there are no configurations which are non-zero and 
$\ol{F(N_k')^{2n+1}} =0$. Hence, also the $2n+1$-point truncated function vanishes, 
and by induction, all odd truncated functions vanish.
\begin{flushright}
$\square$
\end{flushright}

\begin{lemma}\label{even-tf}
All even truncated functions can be written as the sum over all configurations 
with two-point functions (as in Fig.~\ref{circ1}) containing only one cycle
connecting all sites.
\end{lemma}
{\bf Proof}

\begin{figure}[h]
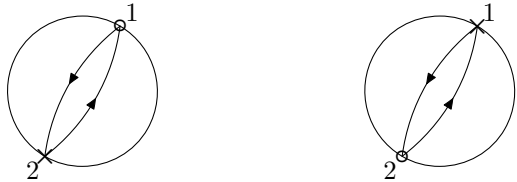

\begin{center}
  \includegraphics{circ-fig2.1}
  \hspace{0.15\textwidth}
  \includegraphics{circ-fig3.1}
\end{center}  
\caption{The non-zero diagrams for $\ol{F(N_k')^2}$}
\label{circ2}
\end{figure}

Consider first the  $2$-point truncated function
\begin{eqnarray}\nonumber
\ol{F(N_k'),F(N_k')}_T & = & \ol{F(N_k')^2} - \ol{F(N_k')}\ol{F(N_k')} \\
& = & \ol{F(N_k')^2} \nonumber \\
& = &  L^{-\nu} \frac{1}{4}\sum_{p_1,p_2}\ol{b(p_1)b^\dagger(p_2)}
\ol{b^\dagger(p_1+ k)b(p_2 +k)} \nonumber
\\ && \qquad \qquad \quad \label{2point-tf}
+\ \ol{b(p_1+k)b^\dagger(p_2+k)}\ol{b^\dagger(p_1)b(p_2)}.
\end{eqnarray}
Using the diagrammatic representation of these functions (as in Fig.~\ref{circ1}),
we see that $\ol{F(N_k'),F(N_k')}_T$ can be written as the sum over all 
(non-zero) diagrams with two sites containing only one cycle, 
cfr.\ (Fig.~\ref{circ2}). 

Consider now the $2n$-point truncated function
\begin{equation}\label{2n-tf}
\omega_L(F(N_k'),\ldots, F(N_k'))_T = \omega_L(F(N_k')^{2n}) -  
\sum_\mathcal{\tau \in P'}\prod_{J \in \tau} \omega_L(F(N_k'),\ldots,F(N_k'))_T,
\end{equation}
and suppose that all $2m$-point functions, with $m < n$ are of the prescribed
form. Furthermore all odd truncated functions vanish (lemma \ref{odd-tf}), 
hence the sum in the second term is a sum over  products of diagrams with one
cycle and an even number of sites.

The first term $\ol{F(N_k')^{2n}}$ can be rewritten in terms of diagrams
(as in Fig.~\ref{circ1}).
We use the fact that diagrams containing several cycles can be written as 
the product of diagrams over less points containing only one cycle, 
because all cycles are independent.
Take a certain partition $\tau \in \mathcal{P}$  of the $2n$ 
sites into subsets of $|J_1|,|J_2|,\ldots,|J_{\tau_{max}}|$ sites, and consider  
now all possible arrow-diagrams where there is for every subset $J_1,J_2,\ldots
\in \tau$ just one cycle, connecting all the points in that subset.
By distributivity we can rewrite this as the product over all subsets $J_i$ of 
the sum of all possible configurations with one cycle on a diagram with 
$|J_i|$ elements.

Since all cycles on diagrams with an odd number of points vanish, we only 
have to deal with the partitions $\tau$ in subsets all containing an even 
number of points. 
In the case of a partition $\tau$ in two or more subsets, we can apply the
induction hypothesis, \ie suppose that the sum over all possible 
diagrams with $2m < 2n$ points with only one cycle is equal to the $2m$-point 
truncated function.
This yields that the term corresponding to the partition $\tau$ can be 
written as the following product
\begin{equation}
\prod_{j_i \in \tau} \ol{F(N_k'),\ldots,F(N_k')}_T.
\end{equation}    
Hence, the $2n$-point truncated functions (\ref{2n-tf}) can be written as  
\begin{eqnarray*}
\ol{F(N_k'),\ldots, F(N_k')}_T &=& \sum_{\text{diagrams with one $2n$-cycle}} +  
\sum_\mathcal{P'}\prod \omega_L(F(N_k'),\ldots,F(N_k'))_T 
\\&& \qquad \qquad  \quad -\ 
\sum_\mathcal{P'}\prod \omega_L(F(N_k'),\ldots,F(N_k'))_T
\\ & = & \sum_{\text{diagrams with one $2n$-cycle}}
\end{eqnarray*}
Also the $2n$-point truncated function is of the prescribed form and by
induction, this is valid for all even truncated functions.

\begin{flushright}
$\square$
\end{flushright}


Now we are ready to calculate the distributions of the quasi-particle density
fluctuations.
\begin{theorem}\label{bkn}
The $k \ne 0$ quasi-particle density fluctuations are 
\begin{itemize}
\item Gaussian and normal 
  \begin{equation}\label{g&n}
    \varphi(F(N_k'))(t) = \exp ( -t^2\zeta(z)/4 )
  \end{equation}
 with 
 \begin{equation}\label{zetaz}
  \zeta(z) = \lambda^{-\nu}\int \!\! \mathrm{d}x \left(\Big( \frac{z\e^{-x^2}}{1 - z\e^{-x^2}}
  \Big)^2 + 
  \frac{z\e^{-x^2}}{1 - z\e^{-x^2}} \right), \qquad z = \e^{\beta\mu},\ \lambda
  = \pi\sqrt{\beta},
 \end{equation} 
 in the normal phase, and in the critical and condensed
 phases ($\mu_L  = \epsilon_L(0) -cL^{-\alpha}$, cfr.\ (\ref{mul-series})) for $k \not\in \{0,1\}^\nu$, or
 for $k \in \{0,1\}^\nu\backslash\{0\}$ and $\alpha < \nu/2 $.
\item Non-Gaussian and normal
 \begin{equation}\label{ng&n}
    \varphi(F(N_k'))(t) = \e^{ -t^2 \zeta(z)/4}\left(\frac{1}{1 + \big(\frac{t}
    {2\beta c}\big)^2}\right)^{\sigma(k)}
 \end{equation}
 with $\zeta(z_*)$ as in (\ref{zetaz}), $z_* = \e^{-\beta\nu\sigma^2}$, and
 $\sigma(k) = 2^{(\nu -k^2)}$, for $k \in \{0,1\}^\nu\backslash\{0\}$ 
 and $\alpha = \nu/2 $.
\item Non-Gaussian and abnormal ($\delta = \alpha - \nu/2$) 
 \begin{equation}\label{ng&a}
   \varphi(F(N_k'))(t) =\left(\frac{1}{1 + \big(\frac{ t}{2\beta
   c}\big)^2}\right)^{\sigma(k)},
 \end{equation}
 with $\sigma(k) = 2^{(\nu -k^2)}$,
 for $k \in \{0,1\}^\nu\backslash\{0\}$ and $\alpha > \nu/2 $.
\end{itemize}
\end{theorem}
{\bf Proof}

Applying the expansion (\ref{charexp}) to the functions $F(N_k')$ yields
the following expression for the characteristic function
\begin{equation}\label{qd-distr}
\ol{\e^{\ii t F(N_k')}} = \exp  \sum_{n \geq 1} \frac{(\ii t)^n}{n!}
\ol{F(N_k'),\ldots,F(N_k')}_T 
\end{equation}
where $\ol{F(N_k'),\ldots,F(N_k')}_T$ are the $n$-point truncated functions.

{\bf Normal Phase:}
Lemma \ref{odd-tf} and lemma \ref{even-tf} about the
truncated functions learn that all odd truncated functions vanish, and that all
$2n$-point truncated functions, $2n>2$ are vanishing, this last fact is easily
seen as follows. From lemma \ref{even-tf}, the $2n$-point truncated functions 
could be written as a the sum over all possible cycles with $2n$ points. 
Such a terms are of the form
\begin{equation}\label{2ncycle}
\frac{L^{-n\nu}}{4^n}\sum_p \ol{b^\dagger(p)b(p)}\ol{b(p +
2k)b^\dagger(p + 2k)} \cdots
\end{equation}
with only one summation index, and $2n$ two-point functions, 
$\ol{b(p +jk)b^\dagger(p + jk)}$, $j = 0,1,2,\ldots$ or with the operators 
in different order. 
All these factors are bounded in the normal phase, and using the techniques of 
lemma \ref{lemma2} it is easy to see that the sum
\begin{equation*}
L^{-\nu} \sum_p \ol{b^\dagger(p)b(p)}
\ol{b(p+k)b^\dagger(p + k)} \cdots \ol{b^\dagger(p+ 3k)b(p + 3k)}
\end{equation*}
converges to a finite integral. 
But we have an extra scaling factor
$L^{-(n-1)/\nu}$, which makes the $2n$-point truncated functions, $2n>2$  
behave like $\order{L^{-(n-1)/\nu}}$. 

Hence, the only extensive term is the contribution of the two-point functions 
(\ref{2point-tf}):
\begin{eqnarray}\nonumber
\lim_{L\to\infty}\ol{F(N_k'),F(N_k')}_T & = & \lim_{L\to\infty} L^{-\nu} 
\frac{\lambda^{-\nu}}{4}\sum_{p_1,p_2}\ol{b(p_1)b^\dagger(p_2)}\ol{b^\dagger(p_1
+ k)b(p_2 +k)} \nonumber
\\ && \qquad \qquad \quad \nonumber
+\ \ol{b(p_1+k)b^\dagger(p_2+k)}\ol{b^\dagger(p_1)b(p_2)}
\\ \nonumber
& = & \frac{1}{2}\int \!\! \mathrm{d}x \left(
\Big(\frac{z\e^{-x^2}}{1 - z\e^{-x^2}} \Big)^2 + 
\frac{z\e^{-x^2}}{1 - z\e^{-x^2}} \right)
\\ \label{2pt-int} & = & \zeta(z)/2,
\end{eqnarray}
with $0 < z < \e^{-\beta\nu\sigma^2}$ for the normal phase and $\lambda = \pi
\sqrt{\beta}$, the thermal wavelength.

One can check that there exists a constant $M(z) > 0$ such that the $2n$-point
truncated functions, are bounded by
\begin{equation}\label{errors}
\frac{2n!}{n! n!}(2n -1)!\ L^{-\nu(n-1)}M(z)^n,
\end{equation}
and  
\begin{equation*}
\lim_{L \to \infty}   \sum_{n \geq 2} \frac{(\ii t)^{2n}}{2n!}
\frac{2n!}{n! n!}(2n -1)!\ L^{-\nu(n-1)}M(z)^n
\end{equation*}
converges to zero as $\order{L^{-\nu}}$. Hence, the error series can be
controlled and the distribution function converges to (\ref{g&n}).

{\bf Critical and Condensed Phase:}
In these phases, the contributions of terms like $\ol{b(q)b^\dagger(q)}$
diverge if $q \in \{0,1\}^\nu$. The rate of divergence depends on the scaling
exponent (\ref{mul-series}), we have
\begin{equation}\label{divergingfactors}
\ol{b^\dagger(q)b(q)} = \frac{1}{\beta c}L^{\alpha} + \order{1} = 
\ol{b(q)b^\dagger(q)}, \qquad \forall
q \in \{0,1\}^\nu.
\end{equation} 

A $2n$-point truncated function consists of terms of the form (\ref{2ncycle}),
with one summation index $p$. Only the first $2^\nu$ terms in this sum over 
$p \in \N$, with $p \in \{0,1\}^\nu$, in such a term contain diverging factors
(\ref{divergingfactors}).

If $k \not\in \{0,1\}^\nu$, the only diverging factors are of the form 
$\ol{b(p)b^\dagger(p)}$ or $\ol{b^\dagger(p)b(p)}$, $p \in \{0,1\}^\nu$. All 
factors with indices $p + jk$, with $j =1,2,\ldots$ and arbitrary $p$, are 
finite. 

A term represented by a cycle which connects alternating symbols $\times$ and
$\circ$, contains $n$ factors of the form  $\ol{b(p)b^\dagger(p)}$ or 
$\ol{b^\dagger(p)b(p)}$ and $n$ factors of the form $\ol{b(p+k)b^\dagger(p+k)}$ or 
$\ol{b^\dagger(p+k)b(p+k)}$. In cycles with arrows connecting two symbols
of the same kind $\times$ or $\circ$, some of these factors are replaced by 
factors $\ol{b(p+jk)b^\dagger(p+jk)}$ with $j =2,3,\ldots$ depending on the
number and the order of such an arrows. 
Hence, the maximal number of diverging factors in a term of the
$2n$-point truncated function is $n$ if $k \not\in \{0,1\}^\nu$, and the terms
containing the largest diverging factors are of order
\begin{equation*}
\order{L^{-n\nu + n\alpha}}.
\end{equation*} 
Since $0 < \alpha \leq \nu/2 -\gamma < \nu$, these terms are vanishing, and the 
$k$-mode quasi-particle density fluctuations, with $k \not\in \{0,1\}^\nu$ are 
normal (no extra scaling exponent $\delta \ne 0$ needed) and Gaussian as 
in the normal phase 
\begin{equation*}
\lim_{L\to\infty}\ol{\e^{\ii t F(N_k')}} =  \e^{-t^2\zeta(z_*)/4}
\end{equation*}
with $\zeta(z_*)$ as in (\ref{zetaz}) and $z_* = \e^{-\beta\nu\sigma^2}$.
  
The situation is different for the $k$-mode fluctuations, with 
$k \in \{0,1\}^\nu$.
Now not only factors with index $p$ are diverging but there are also 
diverging factors with index $p+k \in \{0,1\}^\nu$. The factors with 
indices $p+jk$,
with $j =2,3,4,\ldots$ which appear in cycles where arrows
between symbols of the same type are present, are bounded since 
$jk \in \{0,j\}^\nu$, and thus $p+jk \not\in \{0,1\}^\nu$.  The diagrams with 
the highest number of diverging factors are
those where alternating symbols $\times$ and $\circ$ are connected.  They have
then $2n$ diverging factors of order $\order{L^{\alpha}}$. 
Such terms become extensive if $\alpha \geq \nu/2$.
For a $2n$-point truncated function there are $(2n)!/(n!)^2$ diagrams with
an equal number of both symbols. For such diagrams there are  
$n!(n-1)!$ possible cycles which connect all points alternating the two types 
of symbols.
 Hence there are $(2n)!/n$ terms of leading order. In leading order they are
all equal to
\begin{equation*}
\frac{L^{-n\nu -2n\delta}}{4^n}\sum_{p,p+k \in \{0,1\}^\nu} \Big(\frac{1}{\beta
c}L^{\alpha}\Big)^{2n} = \sigma(k) \Big(\frac{1}{2\beta
c}\Big)^{2n}
\end{equation*}
The sum  $\sum_{p,p+k \in \{0,1\}^\nu}$ yields only a numerical factor 
$\sigma(k) = 2^{(\nu - k^2)}$. 

Hence, inserting this  in expression (\ref{qd-distr}) yields
\begin{eqnarray*}
\lim_{L\to\infty}\ol{\e^{\ii t F(N_k')}} & = & 
\lim_{L\to\infty}
\exp \left( \sum_{n \geq 1} \frac{(\ii t)^n}{n!}
\ol{F(N_k'),\ldots,F(N_k')}_T \right)
\\ &=&
\exp \left( \sigma(k)\sum_{n \geq 1} \frac{(\ii t)^{2n}}{2n!} \frac{2n!}{n}(\frac{1}
{2\beta c})^{2n} \right)
\\ &=&\exp \left( \sigma(k)\sum_{n \geq 1} \frac{1}{n} \left(\frac{- t^2}{(2\beta c)^2}
\right)^n\right)
\\ &=& \left(\frac{1}{1 + \big(\frac{ t}{2\beta c}\big)^2}\right)^{\sigma(k)}.
\end{eqnarray*}
In the case $\alpha =\nu/2$, these terms are extensive as well as the extensive
term from the case  $\alpha <\nu/2$ and the distribution is equal to the
product of the distribution in the two other cases.
\begin{equation*}
\lim_{L\to\infty}\ol{\e^{\ii t F(N_k')}} = \left(\frac{1}{1 + \big(\frac{ t}
{2\beta c}\big)^2}\right)^{\sigma(k)} \e^{-t^2\zeta(z_*)/4}
\end{equation*}
with $\zeta(z_*)$ as in (\ref{zetaz}).
This distribution is normal ($\delta = 0$), but non-Gaussian.
Finally, remark that also in the critical and condensed phases the series over
the subdominant terms can be controlled using similar arguments as in
(\ref{errors}).
\begin{flushright}
$\square$
\end{flushright}

%
%
\subsubsection{Unmodulated quasi-particle density fluctuations}

The situation is similar for $k=0$ quasi-particle fluctuations:
\begin{equation}\label{FN'}
F(N_0') = L^{-\nu/2} \sum_{p \in \N^\nu} b^\dagger(p)b(p) -\ol{ b^\dagger(p)b(p)}
\end{equation}

In order to calculate the characteristic functions, we take again the 
expansion in terms of truncated correlation functions (\ref{charexp}),
\begin{equation*}
  \e^{\ii t F(N_0')} = \exp \sum_{n=1}^\infty \frac{(\ii t)^n}{n!}\ol{F(N_0'),
  \ldots,F(N_0')}_T,
\end{equation*}
and we look for an expression of the $n$-point correlation functions.

\begin{figure}[h]
\begin{center}
   \includegraphics{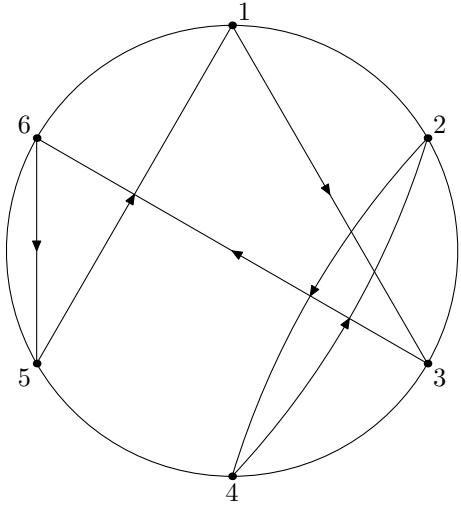}
\end{center}
\caption{A $6$-point diagram with a directed graph consisting of a $2$-point 
cycle and a $4$-point cycle.}\label{circ-0}
\end{figure}

First we need to understand the form of the expectation value of powers of
$F(N_0')$. 
By the quasi-freeness and the gauge invariance of the states, 
the monome $\ol{F(N_0')^n}$ can be decomposed into
a sum over all pair-partitions, where every pair-partition
corresponds to a term consisting of a product of two-point functions. 
Just as for the $k$-mode fluctuations, we can visualise this in a 
diagrammatic representation of $\ol{F(N_0')^n}$, (\eg Fig.~\ref{circ-0}).

Again we draw $n$ sites on a circle, and label them from $1$ to $n$. Each site 
corresponds to 
a factor $L^{-\nu/2}\sum_{p_i} b^\dagger(p_i)b(p_i)- \ol{ b^\dagger(p_i)b(p_i)}$.
Each pair-partition can be represented by a directed graph 
(Fig.~\ref{circ-0}). In every site starts an arrow and ends an arrow, 
every arrow represents a two-point function, this two-point function is 
constructed in the following way, take the creation operator from the starting point 
of the arrow, and combine it with the annihilation operator from 
the endpoint, the order of the operators in the two-point function is 
imposed by the order of the sites, the operator with the lowest site number 
comes first. Loops are now not permitted, the effect of the subtractions
of the expectation values $\ol{ b^\dagger(p_i)b(p_i)}$ in the definition of the
fluctuation observables (\ref{FN'}), lies just in the cancellation of the terms
or graphs containing loops, \ie factors $\ol{ b^\dagger(p_i)b(p_i)}$. 
Every graph consists of a bunch of independent connected subsets or cycles. 
Since every two-point function is only different from zero if the indices of 
both operators are equal, all summation indices of the points in a cycle are equal, and 
there is only one effective summation index for every cycle. 

\begin{lemma}\label{0-n-pt-tf}
$n$-point truncated functions $\ol{F(N_0'),\ldots,F(N_0')}_T$, with $n > 1$, 
can be written as the sum over all diagrams over $n$ points with one cycle 
connecting all $n$ points in the diagram.
\end{lemma}
First note that the one-point truncated function is zero,
\begin{equation*}
\ol{F(N_0')}_T = \ol{F(N_0')} = 0.
\end{equation*}

The two point function reads

\begin{eqnarray*}
\ol{F(N_0'),F(N_0')}_T & = & \ol{F(N_0')^2} - \ol{F(N_0')}\ol{F(N_0')} \\
& = & L^{-\nu}\sum_{p_1,p_2} \ol{b^\dagger(p_1)b(p_2)}\ol{b(p_1)b^\dagger(p_2)}.
\end{eqnarray*}
This is the term represented by the only diagram over two points with one cycle
connecting both points. 

The rest of the proof uses induction. Consider the $n$-point truncated
function and suppose that all $m$-point truncated functions, with $m < n$ 
are of the prescribed form. The $n$-point truncated function is defined as
\begin{equation}\label{0-npt-tf}
\ol{F(N_0'),\ldots,F(N_0')}_T = \ol{F(N_0')^n} - \sum_{\tau \in \mathcal{P'}}
\prod_{J_i \in \tau}\ol{F(N_0'),\ldots,F(N_0')}_T,
\end{equation}
where the sum over $\mathcal{P'}$ is over all partitions $\tau$ into two or
more ordered subsets $J_1,J_2,\ldots \in \tau$. 
Since the one-point truncated function is
zero, partitions containing singletons may also be omitted. 
The first term is the expectation value of the $n^{\text{th}}$ power of
$F(N_0')$. It can be written in terms of diagrams. Using similar arguments as 
in lemma \ref{even-tf} and the induction hypothesis, one can write it as
\begin{equation*}
\ol{F(N_0')^n} = \sum_{\text{$n$-point diagrams with one cycle}} + 
\sum_{\tau \in \mathcal{P'}}\prod_{J_i \in \tau}\ol{F(N_0'),\ldots,F(N_0')}_T,
\end{equation*}
as the second term in this equation is equal to the second term in
(\ref{0-npt-tf}), we conclude that
\begin{equation*}
\ol{F(N_0'),\ldots,F(N_0')}_T = \sum_{\text{$n$-point diagrams with one cycle}}
\hspace{-1.5cm}\cdots
\end{equation*}
or the $n$-point truncated function can also be written as a sum over all
possible diagrams with $n$-points connected through one cycle, and by
induction all truncated functions are of this form.

\begin{flushright}
$\square$
\end{flushright}


\begin{theorem}\label{0qd}

$k=0$ quasi-particle density fluctuations (\ref{FN'}) are  
\begin{itemize}
\item Gaussian and normal 
  \begin{equation}\label{g&n-0}
    \varphi(F(N_0'))(t) = \exp ( -t^2\zeta(z)/2 )
  \end{equation}
 with  $\zeta(z)$ as in (\ref{zetaz}),in the normal phase and in the critical 
 or condensed phases (with $\mu_L  = \epsilon_L(0) -cL^{-\alpha}$) 
 if  $\alpha < \nu/2 $;
\item Non-Gaussian and normal
 \begin{equation}\label{ng&n-0}
    \varphi(F(N_0'))(t) = \e^{-t^2\zeta(z_*)/2}\left(\frac{\e^{-\ii t/\beta c}}
    {1 -  \ii t/\beta c}\right)^{2^\nu}
 \end{equation}
 with $\zeta(z_*)$ as in (\ref{zetaz}), in the critical phase if  
 $\alpha = \nu/2 $;
\item Non-Gaussian and abnormal ($\delta = \alpha - \nu/2$) 
 \begin{equation}\label{ng&a-0}
   \varphi(F(N_0'))(t) = \left(\frac{\e^{-\ii t/\beta c }}
    {1 -  \ii t/\beta c}\right)^{2^\nu}
 \end{equation}
 in the critical or condensed phase if  $\alpha > \nu/2 $.
\end{itemize}
\end{theorem}

{\bf Proof}

In lemma \ref{0-n-pt-tf} we learned that an $n$-point truncated function 
can be written as a sum of terms of the following form
\begin{equation*}
L^{-n\nu/2}\sum_{p\in\N^\nu} \ol{b^\dagger(p)b(p)}\ol{b(p)b^\dagger(p)}\cdots
\end{equation*} 
 where the $n$ factors are either $b^\dagger(p)b(p)$ or $b^\dagger(p)b(p)$. In
 total, there are $(n-1)!$ different cycles connecting all points in a diagram
 of $n$ points, hence there are $(n-1)!$ terms of this form.
We can analyse them as follows
\begin{equation}\label{0-n-pt-decomp}
L^{-n\nu/2}\sum_{p\in\{0,1\}^\nu} \ol{b^\dagger(p)b(p)}\cdots + 
L^{-n\nu/2}\sum_{p\not\in\{0,1\}^\nu} \ol{b^\dagger(p)b(p)}\cdots
\end{equation} 
In the normal phase all factors are bounded, hence since the first term is a
finite sum over bounded terms,  due to the  scaling factor $L^{-n\nu/2}$, 
it vanishes in the thermodynamic limit. 
Using the results of lemma \ref{lemma2}, 
\begin{equation*}
\lim_{L\to \infty}L^{-\nu} \sum_{p\not\in\{0,1\}^\nu}\ol{b^\dagger(p)b(p)} = 
\lambda^{-\nu}\int \!\! \mathrm{d}x\ \frac{z\e^{-x^2}}{1 - z\e^{-x^2}}, 
\end{equation*} 
and since all two-point factors are bounded, also  
the second term in (\ref{0-n-pt-decomp}) is finite. The extra scaling
factors $L^{-n\nu/2}$ make that these terms vanish if $n > 2$. Hence only the
two-point truncated function contains an extensive term, it reads
\begin{eqnarray*}
\lim_{L\to \infty} L^{-\nu} \sum_{p\not\in\{0,1\}^\nu} \ol{b^\dagger(p)b(p)}\ol{b(p)b^\dagger(p)}
& = & \lambda^{-\nu}\int \!\! \mathrm{d}x\ \left(\Big( \frac{z\e^{-x^2}}{1 - z\e^{-x^2}} 
\Big)^2 +   \frac{z\e^{-x^2}}{1 - z\e^{-x^2}} \right)\\
& = & \zeta(z)
\end{eqnarray*} 
with $\zeta(z)$ as in (\ref{zetaz}).

In the critical and condensed phases, the analysis of this second part of
(\ref{0-n-pt-decomp}) is analogous, but now, the first part contains diverging
factors, cfr.\ ($\ref{divergingfactors}$).
If $\alpha < \nu/2$ the scaling exponent $L^{-n\nu/2}$ dominates and the first
term of (\ref{0-n-pt-decomp}) vanishes. The situation is thus similar to the
normal phase, we have normal fluctuations ($\delta =0$) and
the only dominating term is the two-point truncated function,
\begin{equation*}
\lim_{L\to \infty} L^{-\nu} \sum_{p\not\in\{0,1\}^\nu} \ol{b^\dagger(p)b(p)}
\ol{b(p)b^\dagger(p)} = \zeta(z_*), \qquad \text{with}\ z_* =
\e^{-\beta\nu\sigma^2} < 1.
\end{equation*} 

If $\alpha > \nu/2$, the first part in  (\ref{0-n-pt-decomp}) becomes
dominant, the fluctuations become abnormal and non-Gaussian, a scaling factor 
($\delta = \nu/2 -\alpha$) is necessary in order to obtain non-trivial 
distributions. Using the relations (\ref{divergingfactors}), 
the expression for (\ref{0-n-pt-decomp}) is in leading order equal to
\begin{equation*}
L^{-n\nu/2 -n\delta}\sum_{p\in\{0,1\}^\nu} \ol{b^\dagger(p)b(p)}\cdots
= 2^n \frac{1}{(\beta c)^n} + \order{L^{\nu -n\alpha}}.
\end{equation*}  
This yields the following expression for the characteristic function of $F(N_0')$
\begin{eqnarray*}
\exp \sum_{n=1}^\infty \frac{(\ii t)^n}{n!}\ol{F(N_0'),
\ldots,F(N_0')}_T  & = & \exp \sum_{n=2}^\infty \frac{(\ii t)^n}{n!}
(n-1)! 2^\nu \frac{1}{(\beta c)^n} \\
& = & \exp \Big(2^\nu \sum_{n=2}^\infty \left(\frac{\ii t}{\beta c}\right)^n 
\frac{1}{n}  \Big)\\
& = & \left( \frac{\e^{-\ii t/\beta c }}{1 - \ii t/\beta c } \right)^{2^\nu}.
\end{eqnarray*}
As in the case for $k \ne 0$ quasi-particle fluctuations (\ref{errors}), it can
be shown for $k = 0$, that the series over the subdominant contributions vanish
in the infinite volume limit.

If $\alpha = \nu/2$ the dominating terms are both the $\zeta(z_*)$ term in 
the two-point function, and the dominating terms (for $\alpha > \nu/2$) 
in all $n$-point functions. The distribution becomes:
\begin{equation*}
\varphi(F(N_0'))(t) = \e^{ -t^2\zeta(z_*)/2}\left(\frac{\e^{-\ii t/\beta c}}
    {1 -  \ii t/\beta c}\right)^{2^\nu}.
\end{equation*}
\begin{flushright}
$\square$
\end{flushright}

\subsection{$k \in \N^\nu \backslash \{0,1\}^\nu$ density fluctuations}
Now we turn our attention to the density fluctuations of the bare particles.
The transformation to the quasi-particles (\ref{ba}) has hidden the 
influence of the external field, therefore it plays no role in the distribution
functions of the quasi-particles (theorems \ref{bkn} and \ref{0qd}). Its role
will become visible again in the study of the density fluctuations of
the bare particles.  
First we treat the fluctuations with high modulation $k \not\in\{0,1\}^\nu $
(\ref{FNK}). 


\begin{theorem}\label{high}
The $k \in \N^\nu \backslash \{0,1\}^\nu$ density fluctuations 
(\ref{FNK})
are Gaussian
distributed and normal ($\delta = 0$) in all phases, the distribution functions
are given by:
\begin{itemize}
\item in the normal and critical phases,
\begin{equation*}
\varphi(F(N_k))(t) = \exp(-t^2\zeta(z)/4),
\end{equation*}
with $\zeta(z)$ as in (\ref{zetaz}), $z < e^{-\beta\nu\sigma^2}$ in the normal
phase, and $z = e^{-\beta\nu\sigma^2}$ in the critical phase,
\item in the condensed phase,
\begin{equation*}
\varphi(F(N_k))(t) =\exp\left(-t^2/8\left( 2\zeta(z_*) +
\rho_0\coth(\beta\nu\sigma^2/2) \right)\right).
\end{equation*}
\end{itemize}

\end{theorem}
{\bf Proof}
\\
{\bf Normal Phase:}
In the normal phase $\mu_L \to \mu < \mu_*$, we have the following relation
between $a(0)$ and $b(0)$:
\begin{equation*}
a(0) = b(0) + L^\gamma \e^{\ii\phi} \theta(L)
\end{equation*}
where $\theta(L) =\frac{h}{\epsilon_L(0)-\mu_L}$,  an unimportant constant 
converging to a finite value. The $k$-mode density fluctuations can
be written as the sum over the $k$-mode quasi-particle density fluctuations
(\ref{FNK'}), $F_1 = F(N_k')$ and a field fluctuation $F_0$ 
\begin{eqnarray*}
F_L(N_k) & = & L^{-\nu/2}\frac{1}{2} \sum_p  a^\dagger(p)a(p+k)+ a^\dagger(p+k)a(p)\\
& = & L^{-\nu/2}\frac{1}{2} \sum_p  b^\dagger(p)b(p+k)+ b^\dagger(p+k)b(p) + 
L^{\gamma-\nu/2} \theta(L) \frac{1}{2} \left( b^\dagger(k)\e^{\ii \phi} + b(k)
\e^{-\ii \phi} \right) \\ 
& = & F_1 + F_0.
\end{eqnarray*}
In the normal phase, the fluctuations are completely dominated by the first 
term ($F_1 = F(N_k')$), \ie
\begin{equation}\label{F1+F0=F1}
\lim_{L\to \infty} \ol{\e^{\ii t(F_1 + F_0)} - \e^{\ii t F_1}} = 0.
\end{equation}
This can be seen using the following estimates,
\begin{eqnarray*}
|\ol{\e^{\ii t(F_1 + F_0)} - \e^{\ii t F_1}}| & \leq & \int_0^t \!\! 
\mathrm{d}s \
|\ol{\e^{\ii s F_1}F_0\e^{\ii (t-s)(F_1 + F_0) }}| \\
& \leq & \int_0^t \!\! \mathrm{d}s\ \ol{\e^{\ii s F_1} F_0^2\e^{-\ii s F_1 
}}^{1/2},
\end{eqnarray*}
using the Cauchy-Schwarz inequality in the last line.
The expectation value appearing at the rhs in the last line, can be evaluated 
as follows:
\begin{eqnarray}\nonumber
  \ol{\e^{\ii s F_1} F_0^2\e^{-\ii s F_1  }} & = & \sum_{n =0}^\infty 
  \frac{(\ii s)^n}{n!} \ol{[F_1,F_0^2]_n}\\
\label{evolv-n}
&\leq & \sum_{n =0}^\infty \frac{|s|^n}{n!}|\ol{[F_1,F_0^2]_n}|,
\end{eqnarray}
where $[F_1,F_0^2]_0 = F_0^2$ and $[ F_1,F_0^2]_{n+1} = [F_1,[F_1,F_0^2]_n]$.

We now proceed with an estimate of this series. 
The first term reads
\begin{eqnarray*}
\ol{F_0^2} &=&  L^{2\gamma -\nu}\ol{b^\dagger(k) b^\dagger(k) \e^{-2\ii \phi} +
b^\dagger(k)b(k) + b(k)b^\dagger(k) + b(k)b(k)\e^{2\ii \phi}}\\
&=& L^{2\gamma -\nu}\ol{2b^\dagger(k)b(k) +1}.
\end{eqnarray*}
In order to calculate the higher order terms, remark that we need to keep
track only of the part $b^\dagger(k)b(k) + b(k)b^\dagger(k)  = 2b^\dagger(k)b(k)+ 1, $
the other terms in $F_0^2$ and higher order commutators of them with $F_1$ 
contain always an unequal number of creation and annihilation
operators. Hence, their expectation value is zero and they can be forgotten. 
All contributing terms are of the form $b^\dagger(q)b(p)$ and commutation
of such a term with $F_1$ leads to
\begin{equation*}
[F_1,b^\dagger(q)b(p)] = \frac{L^{-\nu/2}}{2}\left( b^\dagger(q+k)b(p) + 
b^\dagger(q-k)b(p)-b^\dagger(q)b(p+k)- b^\dagger(q)b(p-k)  \right),
\end{equation*}   
\ie we find again (at most) 4 terms of the same structure. 
Using now formula (\ref{oL})
\begin{equation*}
|\ol{b^\dagger(q)b(p)}| \leq \ol{b^\dagger(0)b(0)}, \qquad \forall p,q \in
\N^\nu, 
\end{equation*}
we can estimate the $n^{\text{th}}$ order term by
\begin{equation*}
|\ol{[F_1,F_0^2]_n}| \leq 2 L^{2\gamma -\nu} \frac{1}{2^n}L^{-n\nu/2}
4^n \ol{b^\dagger(0)b(0)}.
\end{equation*}
yielding that the whole series is estimated by
\begin{eqnarray}\nonumber
\sum_{n =0}^\infty \frac{|s|^n}{n!}|\ol{[F_1,F_0^2]_n}| &\leq &
L^{2\gamma -\nu}\left( 2 \sum_{n =0}^\infty \frac{|s|^n}{n!} L^{-n\nu/2}
2^n \ol{b^\dagger(0)b(0)} + 1 \right) \\
\label{hk-n}
& \leq & L^{2\gamma -\nu}\left( 2 \ol{b^\dagger(0)b(0)} \e^{2 s L^{-\nu/2}} + 1
 \right).
\end{eqnarray}
The term in brackets is bounded for sufficiently large $L$. Since
$L^{2\gamma -\nu}$ vanishes to zero, the whole term disappears in the
thermodynamic limit. This proves (\ref{F1+F0=F1}), and the distribution of 
the fluctuations is completely determined by the first term.
Hence the distribution of $F(N_k)$ coincides with the distribution of 
$F_1 = F(N_k')$ (\ref{g&n}),
\begin{equation*}
\lim_{L\to \infty} \ol{\e^{itF(N_k)}} = \exp(-t^2 \zeta(z)/4))
\end{equation*}
with $\zeta(z)$ as in (\ref{zetaz}) and $0 < z < \e^{-\beta\nu\sigma^2}$.


{\bf Critical Phase:}
A similar argument as in the normal phase can be developed in the critical phase
($\mu_L = \epsilon_l(0) - c L^{-\alpha}$, with $0 < \alpha < \nu/2 - \gamma$ and
$c > 0$, some unimportant constant) in order to prove that the $F_1 = F(N_k')$ term 
dominates the fluctuations.

The relation between $a(0)$ and $b(0)$ (\ref{ba}) depends now on the scaling exponent
$\alpha$ (\ref{ac}), it reads
$$a(0) = b(0) + \frac{h}{c} L^{\gamma + \alpha}.$$
This yields the following expression for the density fluctuations
\begin{eqnarray*}
F_L(N_k) & = & L^{-\nu/2}\frac{1}{2} \sum_p  a^\dagger(p)a(p+k)+ a^\dagger(p+k)a(p)\\
& = & L^{-\nu/2 }\frac{1}{2} \sum_p  b^\dagger(p)b(p+k)+ b^\dagger(p+k)b(p) + 
L^{\gamma +\alpha - \nu/2}  \frac{h}{2c} ( b^\dagger(k)\e^{\ii \phi} + b(k)
\e^{-\ii \phi} ) \\
& = & F_1 + F_0.
\end{eqnarray*}
The $F_1$ term outrules the $F_0$ contribution, \ie we prove the
same formula as (\ref{F1+F0=F1}) in the critical phase, consider the following
estimate 
\begin{eqnarray*}
|\ol{\e^{\ii t(F_1 + F_0)} - \e^{\ii t F_1}}| & \leq & \int_0^t \!\! 
\mathrm{d}s \
|\ol{\e^{\ii s F_1}F_0\e^{\ii (t-s)(F_1 + F_0) }}| \\
& \leq & \int_0^t \!\! \mathrm{d}s\ \ol{\e^{\ii s F_1} F_0^2\e^{-\ii s F_1  }}^{1/2}
\end{eqnarray*}
The expectation value  $\ol{\e^{\ii s F_1} F_0^2\e^{-\ii s F_1  }}$ can be
expanded as
\begin{multline}\label{evolv}
\ol{\e^{\ii s F_1} F_0^2\e^{-\ii s F_1  }} = \ol{F_0^2} +  \ii s \ol{[F_1,F_0^2]}
\\
- \int_0^s\!\!\mathrm{d}s_1 \! \int_0^{s_1}\!\!\mathrm{d}s_2\  
\ol{\e^{\ii s_2 F_1} \left[F_1,\left[F_1,F_0^2\right]\right]\e^{-\ii s_2 F_1  }}
\end{multline}
The first term  on the rhs of (\ref{evolv}) is bounded as
\begin{eqnarray*}
\ol{F_0^2} & = & L^{\alpha + \gamma -\nu/2} \ol{b(k)b^\dagger(k)+ 
b^\dagger(k)b(k)} \\
& = &  L^{2\alpha + 2\gamma -\nu}2\frac{\e^{-\beta(\epsilon_L(k) -
\epsilon_L(0) + cL^{-\alpha})}}{1-\e^{-\beta(\epsilon_L(k) -
\epsilon_L(0) + cL^{-\alpha})}} +1,
\end{eqnarray*}
and this goes to zero in the thermodynamic limit since $\ol{b^\dagger(k)b(k)}$
is bounded if  $k \not\in \{0,1\}^\nu$ and $2\alpha +2\gamma -\nu < 0$,
cfr.\ (\ref{ac}).

The second term in (\ref{evolv}) is zero by gauge invariance of the states 
$\omega_L$.

The third term in (\ref{evolv}) can be estimated by a similar procedure as 
was used in the normal phase estimating $\ol{\e^{\ii s F_1} F_0^2
\e^{-\ii s F_1  }}$, cfr. (\ref{evolv-n})--(\ref{hk-n}). 
The final results is
\begin{eqnarray*}
|\ol{\e^{\ii t(F_1 + F_0)} - \e^{\ii t(F_1)}}| & \leq & 
 \int_0^t \!\! 
\mathrm{d}s \ \left(
 \order{L^{2\alpha + 2\gamma -\nu}} + 0 + 
 \order{L^{2\gamma + 3\alpha - 2\nu}}\e^{2sL^{-\nu/2}}
\right)^{1/2}
\\
&\stackrel{L\to \infty}\to  & 0,
\end{eqnarray*}
since $2\alpha + 2\gamma -\nu < 0$ and $\alpha < \nu$.
Hence, (\ref{F1+F0=F1}) holds also in the critical phase and 
the distribution of $F(N_k)$ coincides with the distribution of 
$F_1 = F(N_k')$ (\ref{g&n}),
\begin{equation*}
\lim_{L\to \infty} \ol{\e^{itF(N_k)}} = \exp(-t^2\zeta(z_*)/4)),
\end{equation*}
with $\zeta(z_*)$ as in (\ref{zetaz}) and $z_* = \e^{-\beta\nu\sigma^2}$.


{\bf Condensed Phase:} 
In the condensed phase the relation between $a(0)$ and 
$b(0)$ is: $a(0) = b(0) + L^{\nu/2}\sqrt{\rho_0}$, 
this yields the following for the fluctuation $F(N_k)$
\begin{eqnarray*}
F_L(N_k) & = & L^{-\nu/2}\frac{1}{2} \sum_p  a^\dagger(p)a(p+k)+ a^\dagger(p+k)a(p)\\
& = & L^{-\nu/2}\frac{1}{2} \sum_p  b^\dagger(p)b(p+k)+ b^\dagger(p+k)b(p) + 
\sqrt{\rho_0} \frac{1}{2} ( b^\dagger(k)\e^{\ii \phi} + b(k)
\e^{-\ii \phi} ) \\
& = & F_1 + F_0
\end{eqnarray*}
In this  regime (condensed phase and $k \not\in \{0,1\}^\nu$), both terms are
equally contributing. The fluctuations are normal ($\delta = 0 $) and the variances 
satisfy 
\begin{gather*}
 0< \lim_{L\to\infty} \ol{F_1^2} < \infty \\
 0< \lim_{L\to\infty} \ol{F_0^2} < \infty. 
\end{gather*}
 An argument as in the proof of theorem \ref{bkn} can be
 used here in order to prove that 
\begin{equation}\label{decomp-c}
\lim_{L\to\infty}\ol{\e^{\ii t(F_1 + F_0}} = \lim_{L\to\infty}\ol{\e^{\ii tF_1}}
\lim_{L\to\infty}\ol{\e^{\ii tF_0}}.
\end{equation}
We use the expansion of the characteristic functions in terms of truncated
correlation functions (\ref{charexp}), and prove that all truncated
functions vanish 
\begin{equation*}
 \ol{F_1+F_0,F_1+F_0,\ldots,F_1+F_0}_T  \leq \order{L^{\alpha -\nu}},
\end{equation*}
except for the two-point truncated function.
 
{\bf Odd truncated functions vanish.}
First note that the one-point truncated function vanishes.
Consider then the $2n + 1$-point truncated function,
and suppose that all odd m-point truncated functions with $ m < 2n + 1$ are
zero. The expression for the $2n+1$ truncated function reads then:
\begin{equation}\label{odd-tfx}
 \ol{F_1+F_0,F_1+F_0,\ldots,F_1+F_0}_T = \ol{(F_1+F_0)^{2n+1}}  
\end{equation} 
Use now,  
\begin{equation}\label{f1-f2-decomp}
\begin{split}
& F_1 = \left(L^{-\nu/2}/2\sum_p b^\dagger(p)b(p+k)\right) + 
        \left(L^{-\nu/2}/2\sum_p b^\dagger(p+k)b(p)\right)\\
& F_0 = \left( \sqrt{\rho_0}b^\dagger(k)\e^{\ii \phi}/2\right) + 
        \left( \sqrt{\rho_0}b(k)\e^{-\ii \phi}/2\right).
\end{split}
\end{equation} 
The expansion of $\ol{(F_1+F_0)^{2n+1}}$ contains terms,  leaving
out some constants, of the form 
\begin{equation}\label{monoompje}
  L^{-\nu(r+s)/2}\sum_{p_1,\ldots,p_{r+s}}
  \ol{b^\dagger(p_1)b(p_1+k) \cdots b^\dagger(k)}
\end{equation}
where we have $r$ factors of the form $b^\dagger(p)b(p+k)$, $s$ factors
$b^\dagger(p+k)b(p)$, $t$ factors $b^\dagger(k)$ and $u$ factors $b(k)$, with
$r,s,t,u = 1,2,\ldots,2n+1$ and $r+s+t+u = 2n+1$.
If we calculate the sum of the indices of the creation operators minus the 
sum of the indices of the annihilation operators, \ie
$rk -sk +mk -nk$, and if $r,s,m,n = 0,1,\ldots,2n+1$ and $r + s  +n+ m = 2n +1$,
then 
\begin{equation*}
rk -sk +mk -nk \ne 0.
\end{equation*} 
Since this is different from zero, it implies that the monome (\ref{monoompje}) 
vanishes. If we decompose it into products of two-point functions, all terms 
contain at least one two-point function with unequal indices for creation and
annihilation operator, which is zero, hence the whole expression vanishes.   
By induction, all odd truncated functions (\ref{odd-tfx}) vanish.

{\bf All $2n$-point truncated functions, with $2n > 2$ vanish.}
Consider first the non vanishing term, the two-point function
\begin{eqnarray*}
 \ol{F_1+F_0,F_1+F_0}_T  & = & \ol{(F_1+F_0)^2} \\
 & = & \ol{F_1^2} + \ol{F_0^2} \\
 & = & \ol{F_1,F_1}_T + \ol{F_0,F_0}_T, 
\end{eqnarray*}
where the cross-terms $\ol{F_1F_0}, \ol{F_0F_1}$ vanish since the number of
creation and annihilation operators is not equal in those monomes.

Now consider the $2n$-point truncated function, $2n > 2$, and suppose that all
$2m$-point functions, with $2 < 2m < 2n $ are vanishing. By  this assumption we
write the $2n$-pont truncated function as
\begin{equation*}
\ol{F_1+F_0,F_1+F_0,\ldots}_T = \ol{(F_1+F_0)^{2n}} - 
c_2(2n)\ol{F_1+F_0,F_1+F_0}_T + \order{L^{\alpha-\nu}}
\end{equation*}
where $c_2(2n)= (2n)!/(2^nn!)$ is the number of pair-partitions of a set of 
$2n$ elements.
The first term can be expanded as follows. We use first the relation
$F(N_k) = F_1 + F_0$  and we write $\ol{F(N_k)^{2n}}$ as a sum over terms with
the following structure,
\begin{equation}\label{F1F1F0}
\ol{F_1F_1F_0F_1F_0\cdots},
\end{equation} 
where we have $2n$ factors either being $F_0$ or $F_1$. Using simple
combinatorics, we see that we have in total $2^{2n}$ terms, and $(2n)!/(2n
-l)!l!$ terms with $0 \leq l \leq 2n$ factors $F_0$.
We make now a further expansion of such an expression using decompositions
as in (\ref{f1-f2-decomp}),
we get terms consisting of $r$ factors with $L^{-\nu/2}\sum_p 
b^\dagger(p)b(p+k) $, $s$ factors $L^{-\nu/2}\sum_q 
b^\dagger(q+k)b(q)$, $t$ factors $b^\dagger(k)$
and $u$ factors $b(k)$, with $r,s,t,u = 0,1,\ldots,2n$ and 
$r + s = 2n -l$ and $t + u = l$, \ie we have terms of the form
\begin{equation*}
 2^{-2n} (\rho_0)^{l/2}\e^{\ii\phi(t-u)/2} L^{-\nu(2n -l)/2} \! 
 \sum_{p_1,\ldots,p_r}\sum_{q_1,\ldots,q_s}\ol{b^\dagger(k)b^\dagger(k)
 b^\dagger(q_1+k)b(q_1)b^\dagger(p_1)b(p_1+k)\cdots}.
\end{equation*}
Using similar arguments as in the case of odd truncated functions, we see that 
the only terms which are possibly non-zero are those
where $r = s = n - l/2$ and $t = u = l/2$.
These terms can be written as a sum of terms consisting of products of two-point
functions. The sum runs over all pair-partitions of the $4n -l$ operators
in the monome. Since $k \not\in \{0,1\}^\nu$, such a products of two-point
functions contain at most $(n-l/2)/2$ diverging factors (\ref{divergingfactors}), they
are of order $\order{L^{(\alpha/2 -\nu)(n -l/2)}}$ and
since $\nu > \alpha$, such a terms are subdominant. 
The highest order terms are those with the highest number of independent 
summation indices. The non-zero terms with the highest number of independent 
summation indices have $r = (2n-l)/2$
summation indices left, they can be constructed if one takes combinations 
of the operators
$b^\dagger(k)$ and $b(k)$ in factors $\ol{b^\dagger(k)b(k)}$ or 
$\ol{b(k)b^\dagger(k)}$, and if one combines the operators in 
groups with the same summation index $p_i$ of the form
$b^\dagger(p_i)b(p_i+k)$ into pairs with elements from
a group of operators of the form $b^\dagger(p_j +k)b(p_j)$ and vice versa. 
This can be done for every group of operators since the number of groups of 
both types are equal. It yields factors of the form
\begin{equation}\label{f+f-}
\begin{split}
f_{+}(p_i)= \ol{b^\dagger(p_i+k)b(p_j+k)}\ol{b(p_i)b^\dagger(p_j)}
\delta(p_i-p_j),& \quad
\text{or}\\
f_-(p_i)=
\ol{b^\dagger(p_i)b(p_j)}\ol{b(p_i+k)b^\dagger(p_j+k)}\delta(p_i-p_j).&
\end{split}
\end{equation}   
Terms with less summation indices lead to correction terms
which are at most of order $\order{L^{-\nu}}$, \ie
the highest order terms are
\begin{equation} \label{h-order}
\left(L^{-\nu}/2\right)^{r} \sum_{p_1,\ldots,p_{r}}
f_\pm(p_1)\ldots f_\pm(p_{r})\ol{b^\dagger(k)b(k)} \ol{b(k)b^\dagger(k)}\ldots,
\end{equation}
where, depending on the order of the factors, we have more or less factors
$\ol{b(k)b^\dagger(k)}$ or  $\ol{b^\dagger(k)b(k)}$, and
$f_{+}(p)$ or $f_-(p)$ (\ref{f+f-}).
But by symmetry in the expansion (\ref{f1-f2-decomp}),
there are as many terms in the whole sum starting with
$f_+(p_1)$ as with $f_-(p_1)$ or with $\ol{b^\dagger(k)b(k)}$ and  
$\ol{b(k)b^\dagger(k)}$. This holds for all $n$ factors, so in the end we have
a number of terms which can be written as
\begin{multline*}
\left(L^{-\nu}/2\right)^{n - l/2} \sum_{p_1,\ldots,p_{n-l/2}}
\left(f_+(p_1) + f_-(p_1)\right)/2\ldots \left(f_+(p_{n-l/2}) + f_-(p_{n-l/2})
\right)/2 
\\
\left(\left(\ol{b^\dagger(k)b(k)} + \ol{b(k)b^\dagger(k)}\right)/2\right)^{l/2}
\end{multline*}
and this is equal to
\begin{equation}\label{c-form}
\ol{F_1^2/2}^{n-l/2}\ol{F_0^2/2}^{l/2},
\end{equation}
where $l$ is an even number between zero and $2n$.

How many highest order terms (\ref{c-form}) are there  
in the expression for $\ol{(F_1 + F_0)^{2n}}$ ?
There where $(2n)!/(2n-l)!l!$ terms with $0 \leq l \leq 2n$ factors 
$F_0$, cfr.\ (\ref{F1F1F0}),
those terms can be  written as $(2n)!/((l/2)!(n-l/2)!)^2$ terms which
where non zero with $r = s = n -l/2$ and $t = u =l/2$.  
These terms are a sum of terms consisting of products of
pair-partitions. Each original term yields $(n-l/2)!(l/2)!$ 
terms of leading order (\ref{h-order}). The total number of leading order terms
amounts to $(2n)!/((l/2)!(n-l/2)!)$ terms of the form (\ref{c-form}), \ie
\begin{eqnarray*}
\ol{F(N_k)^{2n}} 
&=& \sum_{l = 0}^n (2n)!/((l/2)!(n-l/2)!\ol{F_1^2/2}^{n-l/2}\ol{F_0^2/2}^{l/2}
+ \order{L^{\alpha -\nu}}\\
&=& (2n)!/2^nn!\left( \ol{F_1^2} +  \ol{F_0^2} \right)^n 
+ \order{L^{\alpha -\nu}}\\
&=& c_2(2n) \ol{F(N_k)^2}^n + \order{L^{\alpha -\nu}}.
\end{eqnarray*}
Hence, the $2n$-point correlation function vanishes in the thermodynamic
limit.
The decomposition (\ref{decomp-c}) is valid and 
the limiting distributions can be written as,
\begin{eqnarray*}
\lim_{L\to\infty} \ol{\e^{\ii tF(N_k)}}  & = & \lim_{L\to\infty} \ol{\e^{\ii 
tF_1}}\ol{\e^{\ii tF_0}}
\\
&=& \lim_{L\to\infty} \ol{\e^{\ii 
tF(N_k')}}\ol{\e^{\ii t \frac{\sqrt{\rho_0}}{\sqrt{2}}F(A^+_k)}}
\\
&=& \exp\left(-t^2/8\left( 2 \zeta(z_*) + \rho_0\coth(\beta\nu\sigma^2/2) \right)\right).
\end{eqnarray*}

\begin{flushright}
$\square$
\end{flushright}

\subsection{$k \in \{0,1\}^\nu \backslash \{0\}$ density fluctuations}
We continue now with the study of the low-lying $k$-mode fluctuations.

\begin{theorem}\label{kl-n}
The $k \in \{0,1\}^\nu \backslash \{0\}$ density fluctuations are Gaussian and
normal in the normal phase, the distribution function is given by:
\begin{equation*}
    \varphi(F(N_k))(t) = \exp(-t^2 \zeta(z)/4), \qquad z < 
    \e^{-\beta\nu\sigma^2},
 \end{equation*}
 with $\zeta(z)$ as in (\ref{zetaz}).
\end{theorem}
{\bf Proof}

The same argument as in the case of the $k \in \N^\nu \backslash \{0,1\}^\nu$ density 
fluctuations (theorem \ref{high}) applies here.
\begin{flushright}
$\square$
\end{flushright}

\begin{theorem} \label{kl-cc}
In the critical and condensed phases, \ie when $\mu_L = \epsilon_L(0) -
cL^{-\alpha}$ with $c >0$ and $0 < \alpha \leq \nu/2 -\gamma$, the distribution
of the  $k \in \{0,1\}^\nu \backslash \{0\}$ density fluctuations are 
different in these regions

\begin{enumerate}[(1)]
\item \label{reg1} Gaussian and normal ($\delta = 0$)  
	\begin{equation*}
	  \varphi(F(N_k))(t) = \exp(-t^2\zeta(z_*)/4),
	\end{equation*}
      with $\zeta(z_*)$ as in (\ref{zetaz}),
      if $\alpha < \nu/2 \ \text{and}\  3\alpha < -2\gamma + \nu$,  
\item \label{reg2} Non-Gaussian and normal ($\delta = 0 $) 
      	\begin{equation*}
	  \varphi(F(N_k))(t) = \e^{-t^2\zeta(z_*)/4}\left(\frac{1}
	  {1 + \big(\frac{ t}{2\beta c}\big)^2}\right)^{\sigma(k)},
	\end{equation*}
	with $\sigma(k)  = 2^{(\nu - k^2)}$,
	if $\alpha = \nu/2 \ \text{and}\  \alpha + 2\gamma  < 0$,
\item \label{reg3} Non-Gaussian and abnormal ($\delta = \alpha - \nu/2 $) 
        \begin{equation*}
   	\varphi(F(N_k))(t) =\left(\frac{1}{1 + \big(\frac{ t}{2\beta
   	c}\big)^2}\right)^{\sigma(k)},
	\end{equation*}      	
	if $\alpha > \nu/2 \ \text{and}\  \alpha + 2\gamma  < 0$,
\item \label{reg4} Gaussian  and abnormal ($\delta = \gamma + 3\alpha/2 - \nu/2$)
	\begin{equation*}
 	  \varphi(F(N_k))(t) = 
	  \exp\left(-t^2\frac{h^2}{4\beta c^3}\right),
	\end{equation*} 
	if $3\alpha > -2\gamma + \nu \ \text{and}\  \alpha + 2\gamma  > 0$,
	this last regime includes the the condensed phase, \ie
        $\alpha = \alpha_* = \nu/2-\gamma$. 
\end{enumerate}
\end{theorem}
{\bf Proof}

In the critical and condensed phase 
($\mu_L = \epsilon_L(0) -cL^{-\alpha}$, $0 < \alpha \leq \alpha_*$, $c>0$),
the relation between $a(0)$ and $b(0)$ reads
$a(0) = b(0) + \frac{h}{c} L^{\gamma + \alpha}$. 
This yields the following expression for $F(N_k)$:
\begin{eqnarray}\nonumber
F_L(N_k) & = & L^{-\nu/2 -\delta}\frac{1}{2} 
\sum_p  a^\dagger(p)a(p+k)+ a^\dagger(p+k)a(p)\\
\nonumber
& = & L^{-\nu/2 -\delta}\frac{1}{2} \sum_p  b^\dagger(p)b(p+k)+ b^\dagger(p+k)b(p) + 
L^{\gamma + \alpha - \nu/2 -\delta}  \frac{h}{2c} ( b^\dagger(k)\e^{\ii \phi} + b(k)
\e^{-\ii \phi} ) \\
& = & F_1 + F_0. \label{F1&F0}
\end{eqnarray}
Note that in the condensed phase ($\alpha = \alpha_*$), we can express
$F_0$ as a function of the condensate density, \ie we can substitute 
$\sqrt{\rho_0}$ for $h/c$.

\begin{figure}[h] 
\begin{center}
\includegraphics{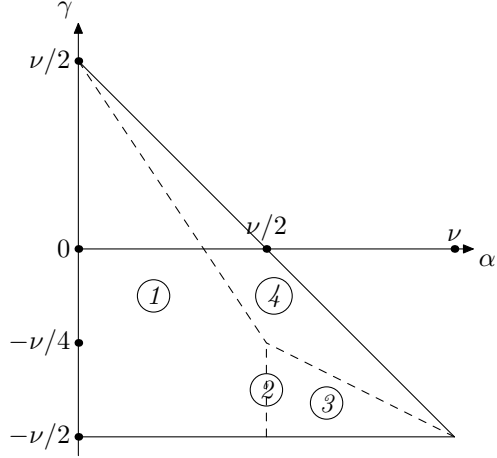}
\end{center}
\caption{Representation of different regions in the 
$(\alpha,\gamma)$-parameterspace}
\label{regions-fig}
\end{figure}

We have to make a distinction between different regions in parameterspace
(\ref{reg1})--(\ref{reg4}), cfr.\ (Fig.~\ref{regions-fig}).


{\bf Region (\ref{reg1})} $\alpha < \nu/2 \ \text{and}\  3\alpha < -2\gamma +
\nu $
\\
In this region the $F_1$ term (\ref{F1&F0}) dominates and the fluctuations can 
be scaled normally, \ie $\delta =0$. The proof is obtained by similar arguments as
in theorem \ref{high} for the case of the normal and critical  phases, we 
have the estimate 
\begin{eqnarray*}
|\ol{\e^{\ii t(F_1 + F_0)} - \e^{\ii tF_1}}| & \leq & \int_0^t \!\! 
\mathrm{d}s \
|\ol{\e^{\ii sF_1}F_0\e^{\ii (t-s)(F_1 + F_0) }}| \\
& \leq & \int_0^t \!\! \mathrm{d}s\ \ol{\e^{\ii s F_1} 
F_0^2\e^{-\ii s F_1  }}^{1/2}.
\end{eqnarray*}
The expectation value on the rhs, can be evaluated 
as follows
\begin{eqnarray}\nonumber
  \ol{\e^{\ii s F_1} F_0^2\e^{-\ii s F_1  }} & = & \sum_{n =0}^\infty 
  \frac{(\ii s)^n}{n!} \ol{[F_1,F_0^2]_n}\\
\label{evolv-cl}
&\leq & \sum_{n =0}^\infty \frac{|s|^n}{n!}|\ol{[F_1,F_0^2]_n}|.
\end{eqnarray}
The last terms of the series are now estimated as before, using 
the following estimate for the $n^{\text{th}}$ order term (see (\ref{hk-n})) 
\begin{equation}\label{n-order-com}
|\ol{[F_1,F_0^2]_n}| \leq 
h^2/2c^2 L^{2\gamma -\nu + 2\alpha} 2^nL^{-n\nu/2}\ol{b^\dagger(0)b(0)}.
\end{equation}
The series is bounded by
\begin{eqnarray} \nonumber
\sum_{n =0}^\infty \frac{|s|^n}{n!}|\ol{[F_1,F_0^2]_n}| &\leq & 
L^{2\gamma -\nu +2\alpha}h^2/4c^2\left( 2 \sum_{n =0}^\infty \frac{|s|^n}{n!} 
L^{-n\nu/2}2^n \ol{b^\dagger(0)b(0)} + 1 \right) 
\\ \nonumber
& \leq & L^{2\gamma -\nu + 2\alpha}h^2/4c^2\left( 2 \ol{b^\dagger(0)b(0)} \e^{2 s L^{-\nu/2}} + 1 \right)
\\ \label{ser-i}
& \leq & \order{L^{2\gamma -\nu + 3\alpha}\e^{2  L^{-\nu/2}}}.
\end{eqnarray}
Since  in region (\ref{reg1}) $3\alpha +2\gamma -\nu < 0$,  
the expectation value  $\ol{\e^{\ii s F_1} F_0^2\e^{-\ii s F_1  }}$ is
vanishing. Hence, the density fluctuations are completely determined by
the first term $F_1 = F(N_k')$ or:
\begin{equation}\label{FN=FN'}
\lim_{L\to\infty} \ol{\e^{\ii tF(N_k)}- \e^{\ii tF(N_k')}} = 0,
\end{equation}
yielding $\varphi (F(N_k))(t) = \varphi (F(N_k'))(t)$, cfr.\ (\ref{g&n}).


{\bf Regions (\ref{reg2})--(\ref{reg3})} $\alpha \geq \nu/2 \ \text{and}\  \alpha + 2\gamma  < 0$
\\
As in region (\ref{reg1}), the $F_1$ term dominates the $F_0$ term in (\ref{F1&F0}), 
the  same argument as used in region (\ref{reg1}) can be used here to prove 
the equality (\ref{FN=FN'}),
the only difference being that other scaling exponents compared to 
expression (\ref{n-order-com}) do appear, \ie
\begin{equation*}
|\ol{[F_1,F_0^2]_n}| \leq 
h^2/2c^2 L^{2\gamma } 2^nL^{-n\alpha}\ol{b^\dagger(0)b(0)}, 
\end{equation*}
substituting these exponents in (\ref{ser-i}) yields
\begin{equation*}
\sum_{n =0}^\infty \frac{|s|^n}{n!}|\ol{[F_1,F_0^2]_n}| \leq  
 \order{L^{2\gamma + \alpha}\e^{2  L^{-\alpha}}}.
\end{equation*}
Since $\alpha + 2\gamma < 0$ in these regions, this term vanishes in the
thermodynamic limit. Hence, also in regions (\ref{reg2})--(\ref{reg3}) we have
that 
\begin{equation*}
\varphi (F(N_k))(t) = \varphi (F(N_k'))(t),
\end{equation*}
and the distribution $\varphi (F(N_k'))$ is given by (\ref{ng&n}) if
$\alpha = \nu/2$, (region \ref{reg2}) or by (\ref{ng&a}) if $\alpha >\nu/2$
(region \ref{reg3}).


{\bf Region (\ref{reg4})} $3\alpha > -2\gamma + \nu $ and $ \alpha + 2\gamma  > 0$ 
\\
In this region the $F_0 $ term  in (\ref{F1&F0}) dominates. 
The variances of $F_0$ are finite, if the scaling exponent is chosen as
$\delta = \gamma + 3\alpha/2 -\nu/2$,
 while the variance of the first term then vanishes.
The distribution is completely dominated by the second term, \ie :
\begin{equation*}
\lim_{L \to \infty} |\ol{\e^{\ii t(F_1 + F_0)} - \e^{\ii t(F_0)}}| = 0.
\end{equation*}
This is proved using similar bounds as before:
\begin{eqnarray*}
|\ol{\e^{\ii t(F_1 + F_0)} - \e^{\ii tF_0}}| & \leq & \int_0^t \! ds \ 
|\ol{\e^{\ii sF_0}F_1\e^{\ii (t-s)(F_1 + F_0) }}| \\
& \leq & \int_0^t \! ds \ \ol{\e^{\ii s F_0} F_1^2\e^{-\ii s F_0  }}^{1/2}.
\end{eqnarray*}
The expectation value appearing is again expanded as:
\begin{equation*}
  \ol{\e^{\ii s F_0} F_1^2\e^{-\ii s F_0  }}  =  \sum_{n =0}^\infty 
  \frac{(\ii s)^n}{n!} \ol{[F_0,F_1^2]_n}.
\end{equation*}
An easy calculation learns that this series is cut off after the third term,
\begin{equation}\label{cf-series}
\ol{\e^{\ii s F_0} F_1^2\e^{-\ii s F_0  }}  = \ol{F_1^2} 
+ \ol{[F_0,F_1^2]}
 + \ol{[F_0,[F_0,F_1^2]]}
\end{equation}
The first term reads
\begin{eqnarray*}
\ol{F_1^2} & = & L^{-\nu - 2\delta} \sum_p\ol{ b^\dagger(p)b(p)} + 
  2\ol{ b^\dagger(p)b(p)}\ol{b^\dagger(p+k)b(p+k)} 
\\ && \qquad \qquad  +\ \ol{ b^\dagger(p+k)b(p+k)} \\
& = & L^{-2\gamma -3\alpha}\sum_{p \in \{0,1\}^\nu} \ldots + 
  L^{-2\gamma -3\alpha}\sum_{p \not\in \{0,1\}^\nu} \ldots
\end{eqnarray*}
The terms in the sum over $ p \in \{0,1\}^\nu$  are at most of order
$$\ol{ b^\dagger(p)b(p)}\ol{ b^\dagger(p+k)b(p+k)} = \order{L^{2\alpha}},$$
and since $2\gamma + 3\alpha > 2\alpha$ in  
region (\ref{reg4}), they are vanishing; 
the second term is a Riemann sum converging to a finite integral,
\begin{equation*}
\lim_{L\to \infty}L^{-\nu}\sum_{p \not\in \{0,1\}^\nu} \ldots = 
\zeta(z_*), 
\end{equation*}
with $\zeta(z_*)$ as in $(\ref{zetaz})$.
The scaling exponent  satisfies $\nu - 2\gamma - 3\alpha > 0$ in region 
(\ref{reg4}), therefore this term vanishes in the  thermodynamic limit as well. 

The second term in (\ref{cf-series}) 
is zero because the constituents are expectation values of 
monomes with an unequal number of creation and annihilation operators
\begin{eqnarray*}
\ii \ol{[F_0,F_1^2]} & = & \ii \ol{F_1[F_0,F_1] +[F_0,F_1]F_1 }
\\ & = &  \frac{h}{2c} L^{ -\gamma - 2 \alpha} \omega_L\!\left(\ii 
   F_1\left(   b(2k)\e^{-\ii\phi}- b^\dagger(2k)\e^{\ii\phi} 
   \right.\right.
\\ & & \qquad \qquad
   +\ \left.\left. 
   b(0)\e^{-\ii\phi}- b^\dagger(0)\e^{\ii\phi} \right) + h.c. \right)
\\ & = & 0.
\end{eqnarray*}

The third term  of (\ref{cf-series})  reads
\begin{equation*}
\ol{[F_0,[F_0,F_1^2]]}  =  \ol{2[F_0,F_1]^2 + F_1 [F_0,[F_0,F_1]] 
+ [F_0,[F_0,F_1]]F_1}.
\end{equation*}
The expectation values of the second and third term in this expression 
vanish since $[F_0,[F_0,F_1]] = 0$, the first term is expanded as  
\begin{eqnarray*}
\ol{[F_0,F_1]^2} & = & -  L^{ - 4\alpha - 2\gamma } \frac{h^2}{4c^2}\left( 
\ol{b^\dagger(0)b(0)} + 1 + \ol{b^\dagger(2k)b(2k)} \right) \\
&\leq &  \order{L^{ -3\alpha -2\gamma}} 
 \end{eqnarray*}
and tend to zero in region (\ref{reg4}).

The  higher order terms in (\ref{cf-series}) vanish since they all 
contain the factors $[F_0,[F_0,F_1]] = 0$.
Hence the distribution of these density fluctuations coincides with the
distribution of the field fluctuations, \ie 
$F_0 = \frac{h}{\sqrt{2}c}F(A^+_k)$ cfr.\ (\ref{cf3}),
\begin{equation*}
\varphi(F(N_k))(t) =
\lim_{L\to\infty}\ol{\e^{\ii tF(N_k)}} = 
\lim_{L\to\infty}\ol{\e^{\frac{\ii th}{\sqrt{2}c}F(A^+_k)}} =
\exp\left(
-t^2\frac{h^2}{4\beta c^3}
\right).
\end{equation*} 
\begin{flushright}
$\square$
\end{flushright}

\subsection{Unmodulated ($k=0$) density fluctuations}

\begin{theorem}\label{theor-k0d}
The limiting distributions of the
$k=0$ density fluctuations (\ref{FN0})
are  
\begin{itemize}
\item Gaussian and normal 
   \begin{equation*}
    \varphi(F(N_0))(t) = \exp ( -t^2\zeta(z)/2 )
   \end{equation*}
   with  $\zeta(z)$ as in (\ref{zetaz}), in the normal phase and in the critical 
   or condensed phases (with $\mu_L  = \epsilon_L(0) -cL^{-\alpha}$) 
   if  $\alpha < \nu/2 $ and if $2\gamma + 3\alpha < \nu$.
\item Non-Gaussian and normal
   \begin{equation*}
      \varphi(F(N_0))(t) = \e^{-t^2\zeta(z_*)/2}\left(\frac{\e^{-\ii t/\beta c}}
      {1 -  \ii t/\beta c}\right)^{2^\nu}
   \end{equation*}
   with $\zeta(z_*)$ as in (\ref{zetaz}), in the critical phase if  
   $\alpha = \nu/2 $ and $\gamma < -\nu/4$,
\item Non-Gaussian and abnormal ($\delta = \alpha - \nu/2$) 
   \begin{equation*}
     \varphi(F(N_0))(t) = \left(\frac{\e^{-\ii t/\beta c }}
     {1 -  \ii t/\beta c}\right)^{2^\nu}
   \end{equation*}
   in the critical phase if  $\alpha > \nu/2 $ and $\alpha <
   -2\gamma$.
\item Gaussian and abnormal ($\delta = \gamma + 3\alpha/2 -\nu/2 $)  
   \begin{equation*}
     \varphi(F(N_0))(t) = \exp \Big( -t^2\frac{h^2}{4\beta c^3}
     \Big),
   \end{equation*} 
   in the critical or condensed phase if  $\alpha > - 2\gamma $ and 
   $\alpha > \nu/3 -2\gamma/3$. In the condensed phase, \ie if $\alpha =
   \alpha_* = \nu/2 - \gamma$, this function can be written in terms of the
   condensate density $\rho_0$, \ie substitute $\rho_0^{3/2}/\beta h $ for
   $h^2/\beta c^3$. 
\end{itemize}
\end{theorem}

{\bf Proof }

The proof goes along the same lines as the proofs of theorem \ref{kl-n} and
theorem \ref{kl-cc}.   
\begin{flushright}
$\square$
\end{flushright}

\section{Concluding Remarks}

Clearly our key contribution consists in the rigorous analysis of the field and
density fluctuations in the Bose gas with attractive boundary conditions. 
Explicit distribution functions of the fluctuations are computed which
are as well of the Gaussian or the non-Gaussian type, of the normal as well of
the abnormal type. These results, clearly indicate the influence on the details 
of the boundary conditions, like the external field (\ref{gammarange}), 
the elasticity of the boundaries (\ref{bound1})--(\ref{bound2}), and the
volume dependence of the chemical potentials (\ref{mul-series}). 

In the technique to study the thermodynamic limit we adopt here, the
interplay between the scaling of the external field (\ref{gammarange}) and the
convergence rate of the chemical potentials (\ref{mul-series}) determines the
limiting thermodynamic phase. This interplay also determines the distribution
and the scaling of the field and density fluctuations. For the field
fluctuations (theorem \ref{fifl} ) and for high-modulated density fluctuations
(theorem \ref{high}), the
structure is determined by the division between normal, critical and
condensed phases. For low-modulated and unmodulated density fluctuations
(theorems \ref{kl-cc} and \ref{theor-k0d}), different regions appear in the 
critical phase where these density fluctuations have different scaling laws and 
distribution functions (cfr.\ Fig.\ \ref{regions-fig}).   
Hence, the study  of the Bose gas with external field, cannot be confined
to Dirichlet boundary conditions or a $\gamma =0$ scaling of the
external field. As our explicit calculations demonstrate, going beyond these
limitations reveals an unexpected richness in the
fluctuation distributions.  This suggests that a choice of a particular
strength of the external field  should be well motivated on physical
grounds.        

The results about the (unmodulated) density  fluctuations 
can be compared with previously obtained results \cite{angelescu:1996}.
There is a correspondence on a heuristic level in the sense that density 
fluctuations are Gaussian and normal in the normal phase, 
non-Gaussian and abnormal in (part of) the critical phase and Gaussian but abnormal 
in the condensed phase, and the form of the explicitly obtained non-Gaussian 
distribution (\ref{ng&a-0}) is of the same structure as the ones calculated by
Angelescu et.\ al.\ in \cite{angelescu:1996}. But the scaling exponents they
calculated are different. They found, in three dimensions respectively
$\delta = 0,0.5,3.5$ for the normal, critical and condensed phases,
whereas we found respectively $\delta =0$, $\delta \in [0,1.5)$, and 
$\delta \in (0,1.5)$, indicating that a difference in the boundary conditions
changes the scaling behaviour as well. Furthermore, we found that
density fluctuations can also be Gaussian or normal or both in the critical 
regime,
depending on different choices of parameters. Sufficiently strongly modulated
fluctuations, \ie if $k \not\in \{0,1\}^\nu$ are always Gaussian and normal,
confirming  that there is a substantial Gaussian element even at or below the 
critical point, cfr.\ the discussion about the classical Curie-Weiss model
in \cite{papangelou:1989,verbeure:1994}. 

The explicit form of the distribution functions can be obtained because of the
quasi-free character of the equilibrium states. This model is original in the
sense that the spectrum shows  an energy gap  in the case of attractive 
boundary conditions. Previous explicit studies of fluctuations where only 
performed in the case of spectra without an energy-gap \cite{wreszinski:1974, ziff:1977, buffet:1983,  fannes:1983, 
nachtergaele:1985, tuyls:1995,broidioi:1996, angelescu:1996, michoel:1999b,
zagrebnov:2001}. 
One would also like to know whether such changes in boundary conditions are as
important in models for interacting Bose gases.

Finally, we want to remark that there are more open problems to be studied in
this model.  In particular we considered here a special thermodynamic limit
using a power low dependence of the chemical potential on the volume, yielding
explicit dependence of the critical exponents on this power law. An other
natural way of taking the thermodynamic limit is a limit by taking the density
constant. It is a question whether in this case the different distribution 
functions can be computed explicitly and whether their behaviour is analogous. 
Our experience by now is that properties of fluctuations are very dependent on 
the type of thermodynamic limit taken. 
This problem is related to the
question about the equivalence of ensembles for the free gas with a finite 
gap in the energy spectrum, this is currently under investigation, and hopefully
sheds new light on this problem. 

\bibliographystyle{/home/lauwers/artikels/myunsrt}
\bibliography{/home/lauwers/artikels/biblio}

\end{document}